\documentclass[journal]{IEEEtran}

\ifCLASSINFOpdf
  \usepackage[pdftex]{graphicx}
  \graphicspath{{./fig/}}
\else
  \usepackage[dvips]{graphicx}
  \graphicspath{{./fig/}}
\fi
\usepackage{pstricks, pstricks-add}
\psset{unit=1mm}
\usepackage{subfig}

\usepackage{amssymb}
\usepackage[cmex10]{amsmath}

\DeclareMathAlphabet{\mathitbf}{OML}{cmm}{b}{it}

\newcommand{\ve}{\mathbf}
\newcommand{\m}{\mathbf}

\newcommand{\mf}[1]{\mathbf{\widetilde{\mathbf{#1}}}} 
\newcommand{\vef}[1]{\mathbf{\tilde{\mathbf{#1}}}} 

\begin{document}
\title{Classical and Bayesian Linear Data Estimators for Unique Word OFDM}

\author{Mario~Huemer,~\IEEEmembership{Senior Member,~IEEE}, Alexander~Onic,~\IEEEmembership{Member,~IEEE} and Christian~Hofbauer,~\IEEEmembership{Member,~IEEE} \\
Alpen-Adria-Universit\"at Klagenfurt, Institute of Networked and Embedded Systems \\
Universit\"atsstr. 65--67, 9020 Klagenfurt, Austria \\
\{mario.huemer, alexander.onic, chris.hofbauer\}@uni-klu.ac.at \\
Phone: +43 463 2700-3660, Fax: +43 463 2700-3679%
\thanks{Christian Hofbauer was funded by the European Regional Development Fund and the Carinthian Economic Promotion Fund (KWF) under grant 20214/15935/23108. He is currently funded by the Austrian Science Fund (FWF): I683-N13.
\copyright 2011 IEEE. Personal use of this material is permitted. Permission from IEEE must be obtained for all other uses, in any current or future media, including reprinting/republishing this material for advertising or promotional purposes, creating new collective works, for resale or redistribution to servers or lists, or reuse of any copyrighted component of this work in other works. DOI: 10.1109/TSP.2011.2164912}}


\maketitle

\begin{abstract}
Unique word -- orthogonal frequency division multiplexing (UW-OFDM) is a novel OFDM signaling concept, where the guard interval is built of a deterministic sequence -- the so-called unique word -- instead of the conventional random cyclic prefix. In contrast to previous attempts with deterministic sequences in the guard interval the addressed UW-OFDM signaling approach introduces correlations between the subcarrier symbols, which can be exploited by the receiver in order to improve the bit error ratio performance. In this paper we develop several linear data estimators specifically designed for UW-OFDM, some based on classical and some based on Bayesian estimation theory. Furthermore, we derive complexity optimized versions of these estimators, and we study their individual complex multiplication count in detail. Finally, we evaluate the estimators' performance for the additive white Gaussian noise channel as well as for selected indoor multipath channel scenarios.
\end{abstract}



%
\IEEEpeerreviewmaketitle

\section{Introduction}

\begin{figure}[!ht]
\footnotesize
\centering
\subfloat[Data structure using CPs]{
\begin{pspicture}(75mm, 12mm)
\psframe(0, 0)(6, 6)
\rput(3, 3){CP1}
\psframe(6, 0)(30, 6)
\psline[linestyle=dashed](24, 0)(24, 6)
\rput(15, 3){Data}
\rput(27, 3){CP1}

\psframe(30, 0)(36, 6)
\rput(33, 3){CP2}
\psframe(36, 0)(60, 6)
\psline[linestyle=dashed](54, 0)(54, 6)
\rput(45, 3){Data}
\rput(57, 3){CP2}

\psframe(60, 0)(66, 6)
\rput(63, 3){CP3}
\psline(66, 0)(75, 0)
\psline(66, 6)(75, 6)
\rput(71, 3){$\cdots$}

\pcline{|<*->|*}(0, 8)(6, 8) \naput{$T_{GI}$}
\pcline{|<*->|*}(6, 8)(30, 8) \naput{$T_{DFT}$}
\pcline{|<*->|*}(36, 8)(60, 8) \naput{$T_{DFT}$}
\end{pspicture} \label{fig:sym_structures_CP}
} \\
\subfloat[Data structure using KSP]{%
\begin{pspicture}(75mm, 6mm)
\psframe(0, 0)(6, 6)
\rput(3, 3){KS}
\psframe(6, 0)(30, 6)
\rput(18, 3){Data}
\psframe(30, 0)(36, 6)
\rput(33, 3){KS}
\psframe(36, 0)(60, 6)
\rput(48, 3){Data}
\psframe(60, 0)(66, 6)
\rput(63, 3){KS}

\psline(66, 0)(75, 0)
\psline(66, 6)(75, 6)
\rput(71, 3){$\cdots$}
\end{pspicture} \label{fig:sym_structures_KSP}
} \\
\subfloat[Data structure using UWs]{%
\begin{pspicture}(75mm, 12mm)
\psframe(0, 0)(6, 6)
\rput(3, 3){UW}
\psframe(6, 0)(24, 6)
\rput(15, 3){Data}
\psframe(24, 0)(30, 6)
\rput(27, 3){UW}

\psframe(30, 0)(48, 6)
\rput(39, 3){Data}
\psframe(48, 0)(54, 6)
\rput(51, 3){UW}

\psline(54, 0)(63, 0)
\psline(54, 6)(63, 6)
\rput(59, 3){$\cdots$}

\pcline{|<*->|*}(0, 8)(6, 8) \naput{$T_{GI}$}
\pcline{|<*->|*}(6, 8)(30, 8) \naput{$T_{DFT}$}
\pcline{|<*->|*}(30, 8)(54, 8) \naput{$T_{DFT}$}
\end{pspicture} \label{fig:sym_structures_UW}
}
\caption{OFDM transmit data structures.}
\label{fig:sym_structures}
\end{figure}
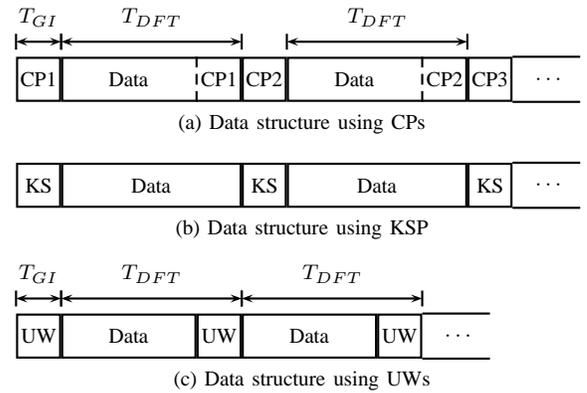

In \cite{Huemer10_1}, \cite{Onic10_1} we introduced an OFDM signaling scheme, where the usual cyclic prefixes (CP) \cite{VanNee00} are replaced by deterministic sequences, that we call unique words (UW). A related but -- when regarded in detail -- also very different scheme is known symbol padded (KSP)-OFDM  \cite{Tang07}--\cite{Welden10}. Fig.~\ref{fig:sym_structures_CP}--\ref{fig:sym_structures_UW} compare the CP-, KSP-, and UW-based OFDM transmit data structures.

In CP- as well as in UW-OFDM the linear convolution of the transmit signal with the channel impulse response is transformed into a cyclic convolution. Note that apart from the very first UW in the symbol stream (see Fig.~\ref{fig:sym_structures}c) each UW plays a double role:  The $(i+1)^\text{th}$ UW represents the tail of the $i^\text{th}$ OFDM symbol, while it additionally represents the `cyclic prefix' for the $(i+1)^\text{th}$ OFDM symbol. However, there are some fundamental differences between the CP-based and the UW-based approach:
\begin{itemize}
\item Different to the CP, the UW is part of the discrete Fourier transform (DFT)-interval as indicated in Fig.~\ref{fig:sym_structures}. Due to that reason the bandwidth efficiencies of UW-OFDM and conventional CP-OFDM are almost identical.
\item The CP is a random sequence, whereas the UW is deterministic. Thus, the UW can optimally be designed for particular needs like synchronization and/or channel estimation purposes at the receiver side.
\end{itemize}

The broadly known KSP-OFDM uses a structure similar to UW-OFDM, since the known symbol (KS) sequence is deterministic as well. The most important difference between KSP- and UW-OFDM is the fact, that the UW is part of the DFT interval, whereas the KS is not. On the one hand this characteristic of the UW implies the cyclic convolution property addressed above, and on the other hand, but least that important, the insertion of the UW within the DFT interval requires to introduce some correlations in the frequency domain, which can advantageously be exploited by the receiver to improve the bit error ratio (BER) performance. Whilst in both schemes the deterministic sequences can be used for synchronization and channel estimation purposes, KSP-OFDM does not feature these correlations. We notice that KSP-OFDM coincides with zero padded (ZP)-OFDM \cite{Lin11} if the KS sequence is set to zero. \smallskip

For single carrier/frequency domain equalization (SC/FDE) systems \cite{Sari94a}--\cite{Reinhardt05}, the benefits of UW based transmission have already sufficiently been studied \cite{Imec00}--\cite{Witschnig02b}, \cite{Huemer03a}--\cite{Witschnig03a}. The introduction of UWs in SC/FDE systems is straightforward, since the data symbols as well as the UW symbols are defined in time domain. In UW-OFDM the data symbols are defined in frequency domain, whereas the UW symbols are defined in time domain, which leads to some difficulties. In \cite{Huemer10_2} we compared the similarities and differences of the UW approach for OFDM and SC/FDE. \smallskip

In our concept described in \cite{Huemer10_1} we suggested to generate UW-OFDM symbols by appropriately loading so-called redundant subcarriers. The minimization of the energy contribution of the redundant subcarriers turned out to be a challenge. We solved the problem by generating a zero UW in a first step, and by adding the desired UW in a separate second step. We showed that this approach generates OFDM symbols with much less redundant energy \cite{Onic10_1} than a single step or direct UW generation approach as e.g.\ described in \cite{Cendrillon01}. Additionally, we optimized the positions of the redundant subcarriers to further reduce their energy contribution. We notice, that the concept in \cite{Cendrillon01} generates
completely different OFDM symbols compared to our approach in \cite{Huemer10_1}, and it has to deal with extremely high symbol energies and with the fact, that the performance depends on the particular shape of the UW. This is clearly in contrast to our approach, where the BER performance is independent of the particular shape of the UW due to the two-step generation approach. The BER behavior only depends on the freely selectable UW energy.

The generation of the zero UW introduces a systematic complex valued block code structure within the sequence of subcarriers. From this point of view, the gain due to the exploitation of correlations in frequency domain can be regarded as a coding gain. Although it seems obvious at first glance to use an algebraic decoding approach, this decoding method fails due to the ill-conditioned nature of the linear system of equations to be solved \cite{Henkel09}, \cite{Hofbauer10_1}.
Instead, we showed that together with a linear minimum mean square error (LMMSE) data estimator (`decoder'), the concept shows a remarkable BER performance, particularly in frequency selective channels, where it clearly outperforms CP-OFDM \cite{Huemer10_1}. The performance can even be increased by allowing some systematic noise in the guard interval \cite{Huemer10_3}. Several other attempts of applying UWs in OFDM systems can be found in the literature, e.g. in \cite{Muck06}--\cite{Jingyi02}. However, in all these approaches the guard interval and thus the UW is not part of the DFT interval. Therefore, in contrast to our UW-OFDM concept described below, no coding is introduced by these approaches.  \smallskip

The aim of this paper is to give a comprehensive view on optimum and suboptimum linear data estimation principles particularly designed and optimized for UW-OFDM. We classify the estimators into classical unbiased estimators and linear Bayesian estimators, respectively. We particularly investigate the theory of the estimators, and we give a comparison in terms of BER performance and in terms of a detailed study of the computational complexities. Furthermore, we emphasize the differences of the derived estimators to their counterparts in competing block oriented approaches like CP-OFDM and SC/FDE. The paper is organized as follows: In Sec.~\ref{sec:uw} we briefly review the procedure of the unique word generation and the overall system model which has already been adressed in \cite{Huemer10_1}, \cite{Onic10_1}, \cite{Huemer10_2}, and \cite{Huemer10_3}. Next we derive data estimators for UW-OFDM using classical estimation theory approaches in Sec.~\ref{sec:zf_estimators} leading to zero forcing (ZF) equalizer 
concepts. In particular we investigate the best linear unbiased estimator (BLUE) which represents a well known concept with a huge number of applications in engineering. In this work it is applied to UW-OFDM for the first time, and it turns out that its construction significantly differs from its counterparts in CP-OFDM and SC/FDE. Different to the BLUE for CP-OFDM and CP-SC/FDE, the determination of the BLUE requires the inversion of a full instead of a diagonal matrix. However, we derive a functionally equivalent but highly complexity reduced version of the BLUE within Sec.~\ref{sec:zf_estimators}. In contrast to CP-OFDM where the BLUE represents the unambiguous zero forcing (ZF) solution, an infinite number of ZF solutions exists for UW-OFDM. We introduce two suboptimum low complexity ZF data estimators, that is the obvious channel inversion (CI) estimator and a quite intuitive estimator that we call time domain windowing (TDW) equalizer. Then, in Sec.~\ref{sec:mmse_estimators} linear Bayesian MMSE 
estimators are regarded. The basic version can already be found in \cite{Huemer10_1}. Similar as for the BLUE we derive a complexity reduced batch solution, and in addition we introduce a highly complexity optimized sequential version of the LMMSE estimator. It turns out that the latter features the lowest complexity of all regarded LMMSE estimator versions. In Sec.~\ref{sec:complexity} we determine and compare the computational complexity of all presented data estimators. Finally, in Sec.~\ref{sec:sim} we highlight the BER performance of the introduced methods in the AWGN channel and in frequency selective indoor multipath environments. We conclude our work in Sec.~\ref{sec:conclusion}.
\bigskip


\noindent
\textit{Notation:} Lower-case bold face variables ($\ve{a},\ve{b}$,\ldots) indicate vectors, and upper-case bold face variables
($\m{A},\m{B}$,\ldots) indicate matrices. To distinguish between time and frequency domain variables, we use a tilde
to express frequency domain vectors and matrices ($\vef{a},\mf{A}$,\ldots), respectively. We further use $\mathbb{R}$ to denote the set
of real numbers, $\mathbb{C}$ to denote the set of complex numbers, $\m{I}$ to denote the identity matrix, $(\cdot)^T$ to denote transposition, $(\cdot)^H$ to denote conjugate transposition, $E[\cdot]$ to denote expectation, and $\mathrm{tr}(\cdot)$ to denote the trace operator. For all signals and systems the usual equivalent complex baseband representation is applied.

\section{Review of UW-OFDM: Unique Word Generation and System Model} \label{sec:uw}

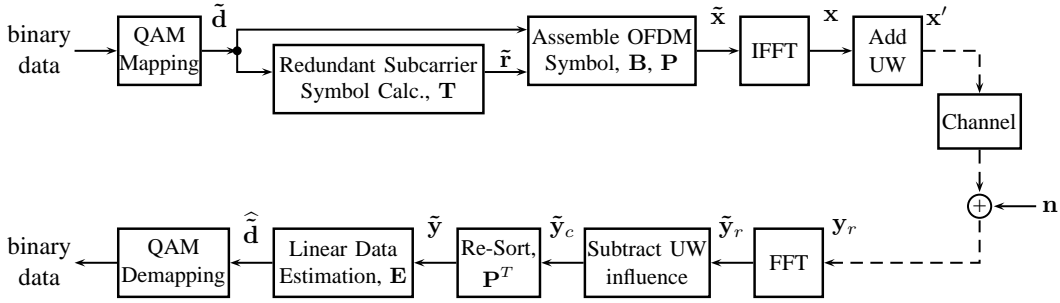
\begin{figure*}[t]
\centering
\begin{pspicture}(140mm, 40mm)
\psset{unit=1mm}
\scalebox{.94}{
\rput(-27, -1.5){
\rput(32, 39){binary}
\rput(32, 35){data}
\psline{->}(37, 37)(43, 37)

\psframe[linewidth=1pt](43, 32)(55, 42)
\rput(49, 39){\small QAM}
\rput(49, 35){\small Mapping}

\psline{->}(55, 37)(60, 37)
\rput(57.5, 42){$\vef{d}$}
\psdot(60, 37)
\psline{->}(60, 37)(60, 34)(65, 34)
\psline{->}(60, 37)(60, 40)(101, 40)

\rput(65, 28){
\psframe[linewidth=1pt](0, 0)(30, 10)
\rput(15, 7){\small Redundant Subcarrier} \rput(15, 3){\small Symbol Calc., $\m{T}$} \rput(33, 8){$\vef{r}$}
}
\rput(0, 30){ \psline{->}(95, 4)(101, 4)

\psframe[linewidth=1pt](101, 2)(125, 12)
\rput(113, 9){\small Assemble OFDM}
\rput(113, 5){\small Symbol, $\m{B}$, $\m{P}$}
\psline{->}(125, 7)(131, 7) \rput(128, 12){$\vef{x}$}

\psframe[linewidth=1pt](131, 2)(141, 12)
\rput(136, 7){\small IFFT}
\psline{->}(141, 7)(147, 7)
\rput(144, 12){$\ve{x}$}

\psframe[linewidth=1pt](147, 2)(157, 12)
\rput(152, 9){\small Add}
\rput(152, 5){\small UW}
\rput(159, 12){$\ve{x}^\prime$}

\psline[linestyle=dashed]{->}(157, 7)(165, 7)(165, 1) }

\rput(159, 23){
\psframe[linewidth=1pt](0, 0)(12, 8)
\rput(6, 4){\small Channel}
}
\cput[framesep=1.5](165, 15){\psdot[dotstyle=+, dotsize=1.5]} \psline[linestyle=dashed]{->}(165, 23)(165, 17) \psline{->}(173, 15)(167, 15) \rput(175, 15){$\ve{n}$}

\rput(32, 9){binary}
\rput(32, 5){data}
\psline{<-}(37, 7)(43, 7)
\psframe[linewidth=1pt](43, 2)(59, 12)
\rput(51, 9){\small QAM}
\rput(51, 5){\small Demapping}

\rput(62, 12){$\widehat{\vef{d}}$}
\psline{<-}(59, 7)(65, 7)

\rput(65, 2){
\psframe[linewidth=1pt](0, 0)(20, 10)
\rput(10, 7){\small Linear Data}
\rput(10, 3){\small Estimation, $\m{E}$} } \psline{<-}(85, 7)(91, 7) \rput(88, 12){$\vef{y}$}

\rput(91, 2){
\psframe[linewidth=1pt](0, 0)(12, 10)
\rput(6, 7){\small Re-Sort,}
\rput(6, 3){\small $\m{P}^T$}
}
\psline{<-}(103, 7)(109, 7)
\rput(106, 12){$\vef{y}_c$}

\rput(109, 2){
\psframe[linewidth=1pt](0, 0)(18, 10)
\rput(9, 7){\small Subtract UW}
\rput(9, 3){\small influence}
}
\psline{<-}(127, 7)(133, 7)
\rput(130, 12){$\vef{y}_r$}

\rput(133, 2){
\psframe[linewidth=1pt](0, 0)(10, 10)
\rput(5, 5){\small FFT}
}
\rput(146, 12){$\ve{y}_r$}

\psline[linestyle=dashed]{<-}(143, 7)(165, 7)(165, 13) }}
\end{pspicture}
\caption{Block diagram of the UW-OFDM transceiver system.}
\label{fig:block_diagram_ISIT}
\end{figure*}

We briefly review our approach of introducing unique words in OFDM time domain symbols, for further details see \cite{Huemer10_1}, \cite{Onic10_1}. A block diagram of the transceiver chain is given in Fig.~\ref{fig:block_diagram_ISIT}. Let $\ve{x}_u\in\mathbb{C}^{N_u \times 1}$ be a predefined sequence which we call unique word. This unique word shall form the tail of each OFDM time domain symbol vector of total length $N$. Hence, an OFDM time domain symbol vector, as the result of a length-$N$-IDFT (inverse DFT), consists of two parts and is of the form $\begin{bmatrix}\ve{x}_d^T & \ve{x}_u^T \end{bmatrix}^T$, whereas only $\ve{x}_d\in\mathbb{C}^{(N-N_u) \times 1}$ is random and affected by the data. In the concept suggested in \cite{Huemer10_1}, \cite{Onic10_1} we generate an OFDM symbol $\ve{x} = \begin{bmatrix}\ve{x}_d^T & \ve{0}^T\end{bmatrix}^T$ with a zero UW in a first step, and we determine the final transmit symbol $\ve{x}' = \ve{x} + \begin{bmatrix}\ve{0}^T & \ve{x}_u^T\end{bmatrix}^T$ by adding 
the desired UW in time domain in a second step. As in conventional OFDM, the quadrature amplitude modulation (QAM) data symbols (denoted by the vector $\vef{d}\in\mathbb{C}^{N_d \times 1}$) and the $N_z$ zero subcarriers (at the band edges and at DC) are specified in frequency domain as part of the vector $\vef{x}$, but here in addition the zero-word is specified in time domain as part of the vector $\ve{x}=\m{F}_N^{-1}\vef{x}$. Here, $\m{F}_N$ denotes the length-$N$-DFT matrix with elements $[\m{F}_N]_{kl}=\mathrm{e}^{-\mathrm{j}\frac{2\pi}{N}kl}$ for $k,l=0,1,...,N-1$. The system of equations $\ve{x}=\m{F}_N^{-1}\vef{x}$ with the introduced features can be fulfilled by spending a set of redundant subcarriers. We let the redundant subcarrier symbols form the vector $\vef{r}\in\mathbb{C}^{N_r \times 1}$ with $N_r=N_u$, we further introduce a permutation matrix $\m{P}\in\mathbb{C}^{(N_d+N_r) \times (N_d+N_r)}$, and form an OFDM symbol (containing $N_z=N-N_d-N_r$ zero subcarriers) in frequency domain by
\begin{equation}
	\vef{x} = \m{B} \m{P} \begin{bmatrix} \vef{d} \\ \vef{r} \end{bmatrix}. \label{equ:2}
\end{equation}
$\m{B}\in\mathbb{C}^{N \times (N_d+N_r)}$ inserts the zero subcarrier symbols, and consists of zero-rows at the positions of the zero subcarriers, and of appropriate unit row vectors at the positions of data and redundant subcarriers. We will detail the reason for the introduction of the permutation matrix $\m{P}$ and its specific construction shortly below. The time -- frequency relation $\m{F}_N^{-1}\vef{x}=\ve{x}$ can now be written as $\m{F}_N^{-1} \m{B} \m{P} \begin{bmatrix} \vef{d}^T & \vef{r}^T \end{bmatrix}^T = \begin{bmatrix}\ve{x}_d^T & \ve{0}^T \end{bmatrix}^T.$
With
\begin{equation}
 		\m{M}=\m{F}_N^{-1} \m{B} \m{P}= \begin{bmatrix} \m{M}_{11} & \m{M}_{12} \\ \m{M}_{21} & 																								\m{M}_{22}\end{bmatrix}, \label{equ:syst019}
\end{equation}
where $\m{M}_{kl}$ are appropriate sized sub-matrices, it follows that $\m{M}_{21} \vef{d} + \m{M}_{22} \vef{r} = \ve{0}$, and hence
$\vef{r} = -\m{M}_{22}^{-1}\m{M}_{21} \vef{d}$. With the matrix
\begin{equation}
	\m{T} = -\m{M}_{22}^{-1}\m{M}_{21} \in\mathbb{C}^{N_r \times N_d}, \label{equ:syst020}
\end{equation}
the vector of redundant subcarrier symbols can thus be determined by the linear mapping
\begin{equation}
\vef{r} = \m{T} \vef{d}, \label{equ:3}
\end{equation}
cf. Fig.~\ref{fig:block_diagram_ISIT}. The construction of $\m{T}$ and thus also the energy of the redundant subcarrier symbols highly depend on the choice of $\m{P}$.
The mean symbol energy $E_{\ve{x}'}=E[\ve{x}'^H\ve{x}']$ can be calculated to
\begin{equation}
		E_{\ve{x}'} = \frac{1}{N} \big(\underbrace{N_d \sigma_d^2}_{E_{\vef{d}}}  + \underbrace{\sigma_d^2 																						\mathrm{tr}(\m{T}\m{T}^H)}_{E_{\vef{r}}}\big) + \underbrace{\ve{x}_u^H\ve{x}_u}_{E_{\ve{x}_u}},
		\label{equ:trans019}
\end{equation}
cf.~\cite{Onic10_1}. $\frac{E_{\vef{d}}}{N}$ and $\frac{E_{\vef{r}}}{N}$ describe the contributions of the data and the redundant subcarrier symbols to the total mean symbol energy before the addition of the UW, respectively, and $E_{\ve{x}_u}$ describes the contribution of the UW. In \cite{Huemer10_1} we suggested to choose $\m{P}$ by a minimization of the symbol energy $E_{\ve{x}'}$ which leads to the optimization problem
\begin{equation}
  \m{P}=\mathrm{argmin}\left\{\mathrm{tr}(\m{T}\m{T}^H)\right\}, \label{equ:uw001}
\end{equation}
where $\m{T}$ is derived from \eqref{equ:syst020} and \eqref{equ:syst019}, respectively. In Sec. \ref{sec:sim} we give an example of the optimum redundant subcarrier distribution for a specific parameter setup.\smallskip

With \eqref{equ:3} the vector $\vef{c}_s = \begin{bmatrix}\vef{d}^T & \vef{r}^T\end{bmatrix}^T$ of data and redundant subcarrier symbols can be written in the form
\begin{equation}
\vef{c}_s = \begin{bmatrix}\vef{d}\\ \vef{r}\end{bmatrix} = \begin{bmatrix}\m{I}\\\m{T}\end{bmatrix} \vef{d} = \m{G}_s\vef{d}.
						\label{equ:uw002}
\end{equation}
In \eqref{equ:uw002} the matrix
\begin{equation}
	\m{G}_s = \begin{bmatrix} \m{I} \\ \m{T} \end{bmatrix} \in\mathbb{C}^{(N_d+N_r)\times N_d} \label{eq:codegenerator}
\end{equation}
can be interpreted as the code generator matrix for a systematic complex valued block code, that generates the code words $\vef{c}_s$. Note, that the subscript '$_s$' in the code words $\vef{c}_s$ and in the code generator matrix $\m{G}_s$ denotes \emph{sorted}. In contrast to \cite{Huemer10_1} and \cite{Onic10_1} we do not incorporate the permutation matrix $\m{P}$ into the code generator matrix in this work. As we will see later on this is essential for deriving low complexity receiver concepts. With \eqref{equ:uw002} and with the frequency domain version of the UW $\vef{x}_u=\m{F}_N \begin{bmatrix} \ve{0}^T&\ve{x}_u^T \end{bmatrix}^T$ the transmit symbol can now also be written as
\begin{equation}
	\ve{x}' = \m{F}_N^{-1} (\m{BPG}_s\vef{d}+\vef{x}_u). \label{eq:tx_generation2}
\end{equation}

\smallskip
After transmission over a dispersive (e.g.\ multipath) channel a received OFDM time domain symbol can be modeled as
\begin{eqnarray}
	\ve{y}_r 	&=& \m{H}_c \ve{x}'+ \ve{n} \\
	&=& \m{H}_c \m{F}_N^{-1} (\m{BPG}_s\vef{d}+\vef{x}_u) + \ve{n},
\end{eqnarray}
where $\ve{n}\in\mathbb{C}^{N \times 1}$ represents a zero-mean Gaussian (time domain) noise vector with the covariance matrix $\sigma_n^2 \m{I}$, and $\m{H}_c\in\mathbb{C}^{N\times N}$ denotes a cyclic convolution matrix originating from the zero-padded vector of channel impulse response coefficients $\ve{h}_c\in\mathbb{C}^{N\times 1}$. After applying a DFT to obtain $\vef{y}_r=\m{F}_N\ve{y}_r$, we exclude the zero subcarriers from further operation, which leads to the down-sized vector $\vef{y}_d=\m{B}^T\vef{y}_r$ with $\vef{y}_d\in\mathbb{C}^{(N_d+N_r) \times 1}$:
\begin{equation}
		\vef{y}_d = \m{B}^T \m{F}_N \m{H}_c \m{F}_N^{-1} (\m{BPG}_s\vef{d} + \vef{x}_u) + \m{B}^T \m{F}_N \ve{n}.
\end{equation}
The matrix $\mf{H}_c=\m{F}_N \m{H}_c \m{F}_N^{-1}$ is diagonal and contains the sampled channel frequency response on its main diagonal. $\mf{H}_d = \m{B}^T \m{F}_N \m{H}_c \m{F}_N^{-1} \m{B}$ with $\mf{H}_d\in\mathbb{C}^{(N_d+N_r) \times (N_d+N_r)}$ is a down-sized version of the latter excluding the entries corresponding to the zero subcarriers. The received symbol can now be written in the form of the \emph{affine} model
\begin{equation}
		\vef{y}_d = \mf{H}_d\m{PG}_s\vef{d} + \mf{H}_d\m{B}^T \vef{x}_u + \m{B}^T \m{F}_N \ve{n}.
\end{equation}
Note that (assuming that the channel matrix $\mf{H}_d$ or at least an estimate of it is available) $\mf{H}_d \m{B}^T \vef{x}_u$ represents the known portion contained in the received vector $\vef{y}_d$ originating from the UW. As a first preparatory step we therefore subtract the UW influence to obtain the corrected symbol in the form of the \emph{linear} model
\begin{eqnarray}
		\vef{y}_c &=& \vef{y}_d - \mf{H}_d \m{B}^T \vef{x}_u \label{equ:system_model_005} \\
		          &=& \mf{H}_d\m{PG}_s\vef{d} + \vef{w}, \label{equ:system_model_004}
\end{eqnarray}
with the noise vector $\vef{w} = \m{B}^T \m{F}_N \ve{n}$. For the low complexity versions of the BLUE and the LMMSE estimator to be derived in the subsequent sections it additionally turns out to be quite advantageous to re-sort the receive vector by applying $\m{P}^T$ to separate the data subcarrier symbols and the redundant subcarrier symbols. The re-sorted receive vector $\vef{y}$ follows to
\begin{align}
	\vef{y} 	&= \m{P}^T \vef{y}_c 					\label{equ:system_model_003} \\
						&= \m{P}^T\mf{H}_d\m{P}\m{G}_s\vef{d} + \m{P}^T\m{B}^T \m{F}_N \ve{n}. 		\label{equ:system_model_002}
\end{align}
With the re-sorted (and still diagonal) channel matrix $\mf{H}_s = \m{P}^T \mf{H}_d\m{P}$ and the noise vector $\vef{v}=\m{P}^T \m{B}^T \m{F}_N \ve{n}$ we finally arrive at the \emph{linear} model
\begin{equation}
	\vef{y} 	= \mf{H}_s\m{G}_s\vef{d} + \vef{v}. \label{equ:system_model_001}
\end{equation}

\section{Classical Data Estimators -- Zero Forcing Solutions}
\label{sec:zf_estimators}
In this section we consider classical unbiased data estimators of the form
\begin{equation}
	\widehat{\vef{d}} = \m{E} \vef{y}, \label{equ:001}
\end{equation}
where $\m{E}\in\mathbb{C}^{N_d \times (N_d+N_r)}$ describes the equalizer. Note that in classical estimation the data vector is assumed to be deterministic but unknown. In order for the estimator to be unbiased we require
\begin{equation}
	E[\widehat{\vef{d}}] = E[\m{E} \vef{y}] = \m{E}\mf{H}_s\m{G}_s\vef{d} = \vef{d}.
\end{equation}
Consequently, the unbiased constraint takes on the form
\begin{equation}
	\m{E}\mf{H}_s\m{G}_s = \m{I}, \label{equ:ZF_006}
\end{equation}
which is equivalent to the ZF criterion for linear equalizers. The solution to \eqref{equ:ZF_006} is ambiguous. To show this we consider a singular value decomposition of $\mf{H}_s\m{G}_s\in\mathbb{C}^{(N_d+N_r)\times N_d}$ as
\begin{equation}
	\mf{H}_s\m{G}_s = \m{U}\begin{bmatrix} \m{\Sigma} \\ \m{0} \end{bmatrix}\m{V}^H, \label{equ:ZF_030}
\end{equation}
with unitary matrices $\m{U}\in\mathbb{C}^{(N_d+N_r)\times (N_d+N_r)}$ and $\m{V}\in\mathbb{C}^{N_d\times N_d}$, and with the diagonal matrix $\m{\Sigma}\in\mathbb{R}^{N_d\times N_d}$ having as its main diagonal the singular values of $\mf{H}_s\m{G}_s$. With \eqref{equ:ZF_030} the unbiased constraint (or ZF criterion) \eqref{equ:ZF_006} becomes
\begin{equation}
	\m{E}\m{U}\begin{bmatrix} \m{\Sigma} \\ \m{0} \end{bmatrix}\m{V}^H = \m{I}. \label{equ:ZF_031}
\end{equation}
It is easy to see that \eqref{equ:ZF_031} and therefore also \eqref{equ:ZF_006} is fulfilled by every equalizer of the form
\begin{equation}
	\m{E} = \m{V}\begin{bmatrix} \m{\Sigma}^{-1} & \m{A} \end{bmatrix}\m{U}^H \label{equ:ZF_032}
\end{equation}
with arbitrary $\m{A}\in\mathbb{C}^{N_d\times N_r}$. We notice that the fact that the ZF solution is ambiguous distinguishes UW-OFDM from competing block oriented single input single output (SISO) approaches like e.g.\ CP-OFDM and CP-SC/FDE. For CP-OFDM the channel inversion receiver $\m{E} = \mf{H}_d^{-1}$ represents the unambiguous ZF solution which also corresponds to the optimum data estimator, cf.~\cite{VanNee00}.
For CP-SC/FDE the ZF solution is also unambiguous, it is given by the inverse of the diagonal symbol spaced channel matrix which contains the influence of the transmit pulse shaping filter, the dispersive (e.g. multipath) channel, and the receiver filter (e.g. a matched filter), cf. \cite{Huemer03b}. \smallskip

Since the solution to the unbiased constraint is not unambiguous it makes sense to look for the optimum solution which is commonly known as the best linear unbiased estimator.

\subsection{Best Linear Unbiased Estimator (BLUE)}
By applying the Gauss-Markov theorem \cite{Kay93} to \eqref{equ:system_model_001}, and with the noise covariance matrix $\m{C}_{\tilde{v}\tilde{v}} = E \left[\vef{v} \vef{v}^H \right]=N \sigma_n^2 \m{I}$, the BLUE and consequently the optimum ZF equalizer follows to
\begin{equation}
	\m{E}_\mathrm{BLUE} = (\m{G}_s^H\mf{H}_s^H \mf{H}_s\m{G}_s)^{-1} \m{G}_s^H\mf{H}_s^H. \label{equ:ZF_008}
\end{equation}
$\m{E}_\mathrm{BLUE}$ as given in \eqref{equ:ZF_008} represents the pseudoinverse of $\mf{H}_s\m{G}_s$. Since the noise in \eqref{equ:system_model_001} is assumed to be Gaussian, \eqref{equ:ZF_008} is also the minimum variance unbiased (MVU) estimator. The covariance matrix of $\widehat{\vef{d}}=\m{E}_\mathrm{BLUE}\vef{y}$, or equivalently the covariance matrix of the error $\vef{e} = \vef{d}-\widehat{\vef{d}}$ is given by
\begin{equation}
	\m{C}_{\tilde{e}\tilde{e}} = N \sigma_n^2 (\m{G}_s^H\mf{H}_s^H \mf{H}_s\m{G}_s)^{-1}. \label{equ:ZF_009}
\end{equation}
With the singular value decomposition as in \eqref{equ:ZF_030}, and after some rearrangements using standard matrix algebra, \eqref{equ:ZF_008} can immediately be re-written as
\begin{equation}
	\m{E}_\mathrm{BLUE} =  \m{V}\begin{bmatrix} \m{\Sigma}^{-1} & \m{0} \end{bmatrix}\m{U}^H.
\end{equation}
By comparing this result with \eqref{equ:ZF_032} it can be concluded that $\m{E}_\mathrm{BLUE}$ corresponds to the solution in \eqref{equ:ZF_032} for the particular case $\m{A}=\m{0}$. $\m{E}_\mathrm{BLUE}$ is in general a full matrix, which is in contrast to CP-OFDM and CP-SC/FDE, where the BLUE is given by a diagonal matrix.

\subsection{Complexity Optimized Version of the BLUE} \label{sec:BLUE_optimized}
One drawback of the BLUE represented as in \eqref{equ:ZF_008} is the fact, that an $N_d\times N_d$ matrix has to be inverted to determine the equalizer. In this section we derive a significantly complexity reduced version of the BLUE by exploiting the simple structures of $\m{G}_s$ and $\mf{H}_s$, respectively. For this purpose we decompose $\mf{H}_s$ as
\begin{equation}
	\mf{H}_s = \begin{bmatrix} \mf{H}_{s,1} & \m{0} \\ \m{0} & \mf{H}_{s,2} \end{bmatrix}, \label{equ:ZF013}
\end{equation}
with the diagonal matrices $\mf{H}_{s,1}\in\mathbb{C}^{N_d \times N_d}$ and $\mf{H}_{s,2}\in\mathbb{C}^{N_r \times N_r}$. With \eqref{eq:codegenerator} it follows that
\begin{equation}
	\mf{H}_s\m{G}_s = \begin{bmatrix} \mf{H}_{s,1} & \m{0} \\ \m{0} & \mf{H}_{s,2} \end{bmatrix} \begin{bmatrix} \m{I} \\ \m{T} \end{bmatrix}               = \begin{bmatrix} \mf{H}_{s,1} \\ \mf{H}_{s,2} \m{T} \end{bmatrix}, \label{equ:compl_004}
\end{equation}
and the expression $(\m{G}_s^H\mf{H}_s^H \mf{H}_s\m{G}_s)^{-1}$ appearing in \eqref{equ:ZF_008} and \eqref{equ:ZF_009} can be written as
\begin{equation}
	(\m{G}_s^H\mf{H}_s^H \mf{H}_s\m{G}_s)^{-1} = (\mf{H}_{s,1}^H \mf{H}_{s,1} +\m{T}^H \mf{H}_{s,2}^H \mf{H}_{s,2} \m{T})^{-1}. \label{equ:compl_001}
\end{equation}
We introduce the real diagonal matrices
\begin{eqnarray}
	\m{D}_1 &=& \mf{H}_{s,1}^H \mf{H}_{s,1}\in\mathbb{C}^{N_d \times N_d}, \\
	\m{D}_2 &=& \mf{H}_{s,2}^H \mf{H}_{s,2}\in\mathbb{C}^{N_r \times N_r},
\end{eqnarray}
and apply the matrix inversion lemma, cf. \cite{Kay93}, to the right hand side of \eqref{equ:compl_001} to obtain
\begin{eqnarray}
	\lefteqn{(\m{G}_s^H\mf{H}_s^H \mf{H}_s\m{G}_s)^{-1} = } \nonumber \\
	      & & \m{D}_1^{-1}-\m{D}_1^{-1}\m{T}^H(\m{T}\m{D}_1^{-1}\m{T}^H+\m{D}_2^{-1})^{-1}\m{T}\m{D}_1^{-1}.  \label{equ:compl_002}
\end{eqnarray}
The inversions of the real diagonal matrices $\m{D}_1$ and $\m{D}_2$ are trivial, and the additional matrix $(\m{T}\m{D}_1^{-1}\m{T}^H+\m{D}_2^{-1})$ to be inverted is Hermitian and only has the dimension $N_r \times N_r$. Furthermore, the expression $\m{T}\m{D}_1^{-1}$ (and its Hermitian transpose) occurs repeatedly in \eqref{equ:compl_002} which allows for further complexity reduction.

In section \ref{sec:complexity} we will study the complexity of the different representations of the BLUE. We note that the derivation of the complexity reduced version of the BLUE has mainly been made possible by the re-sorting (multiplication with $\m{P}^T$) of the data and redundant subcarrier symbols in \eqref{equ:system_model_002}.

\subsection{Sub-Optimum ZF Receiver Structures}
Any unbiased linear data estimator, or equivalently, any linear zero forcing equalizer has to fulfill \eqref{equ:ZF_006}. As already shown above the ZF solution is ambiguous for the UW-OFDM transmission model described in \eqref{equ:system_model_001}. Another quite intuitive and straightforward ZF solution is given by
\begin{equation}
	\m{E}_\mathrm{CI} = \begin{bmatrix} \m{I} & \m{0} \end{bmatrix} \mf{H}_s^{-1}. \label{equ:ZF_003}
\end{equation}
This equalizer inverts the channel $\mf{H}_s$ first, and the data symbols are extracted subsequently. Clearly this procedure fulfills \eqref{equ:ZF_006}. In the following we will refer to this equalizer as the channel inversion (CI) receiver. Using the decomposition of $\mf{H}_s$ as in \eqref{equ:ZF013}, \eqref{equ:ZF_003} can be simplified to
\begin{equation}
	\m{E}_\mathrm{CI} = (\mf{H}_{s,1})^{-1} \begin{bmatrix} \m{I} & \m{0} \end{bmatrix} . \label{equ:ZF_014}
\end{equation}
The channel inversion receiver represents a low complex solution since $\mf{H}_{s,1}$ has a diagonal structure, but it does not take advantage of the correlations introduced by $\m{G}_s$ at the transmitter side. The covariance matrix of $\widehat{\vef{d}}=\m{E}_\mathrm{CI}\vef{y}$, or equivalently the covariance matrix of the error $\vef{e} = \vef{d}-\widehat{\vef{d}}$ can easily shown to be
\begin{equation}
	\m{C}_{\tilde{e}\tilde{e}} = N \sigma_n^2 (\mf{H}_{s,1}^H \mf{H}_{s,1})^{-1}. \label{equ:ZF_010}
\end{equation}

Next we address another quite intuitive equalizer that exploits the a-priori knowledge, that the guard interval samples of an UW-OFDM symbol must be zero after the channel inversion in the noiseless case. In the presence of noise we therefore simply force the guard interval samples to zero which is achieved by an equalizer of the form
\begin{equation}
	\m{E}_\mathrm{TDW} =  \begin{bmatrix} \m{I} & \m{0} \end{bmatrix}\m{P}^T\m{B}^T \m{F}_N \m{W} 																												\m{F}_N^{-1}\m{B}\m{P}\mf{H}_s^{-1}, \label{equ:ZF_004}
\end{equation}
where
\begin{equation}
	\m{W} = \begin{bmatrix} \m{I} & \m{0} \\ \m{0} & \m{0}\end{bmatrix}. \label{equ:ZF_034}
\end{equation}
The time domain windowing (TDW) equalizer starts with an inversion of the channel, next the permutation is applied and the zero subcarrier symbols are added again in order to be able to transform back to time domain with a length-$N$-IDFT. Here a windowing (described by $\m{W}$) takes place, where the guard interval samples are forced to zero. Next a transformation back to frequency domain is performed, the zero subcarriers are excluded again, a re-sorting is done, and finally the data symbols are extracted. It can easily be shown, that $\m{E}_\mathrm{TDW}$ also fulfills \eqref{equ:ZF_006}. Note that the TDW equalizer also represents a quite low complex solution since none of the individual operations requires a full matrix multiplication, in fact most of the steps apart from DFT and IDFT are trivial. The covariance matrix of $\widehat{\vef{d}}=\m{E}_\mathrm{TDW}\vef{y}$, or equivalently the covariance matrix of the error $\vef{e} = \vef{d}-\widehat{\vef{d}}$ is given by
\begin{align}
	\m{C}_{\tilde{e}\tilde{e}} &= N \sigma_n^2 \m{E}_\mathrm{TDW}\m{E}_\mathrm{TDW}^H. \label{equ:ZF_011}
\end{align}

\section{Linear Bayesian Data Estimators -- LMMSE Solutions}
\label{sec:mmse_estimators}
We now turn to the widely used linear minimum mean square error data estimator which is derived with the help of the Bayesian approach. In the Bayesian approach the data vector is assumed to be the realization of a random vector instead of a deterministic and unknown vector as in the classical estimation theory applied above. In the following we derive the LMMSE batch solution, next we formulate a complexity optimized version of the LMMSE batch solution, and finally we derive a highly complexity optimized version of the sequential LMMSE estimator.

\subsection{LMMSE Batch Solution}
By applying the Bayesian Gauss-Markov theorem \cite{Kay93} to \eqref{equ:system_model_001}, where we now assume $\vef{d}$ to be the realization of a random vector, and by using $\m{C}_{\tilde{d}\tilde{d}} = \sigma_d^2 \m{I}$ and $\m{C}_{\tilde{v}\tilde{v}} = N \sigma_n^2 \m{I}$ the LMMSE equalizer follows to
\begin{equation}
		\m{E}_\mathrm{LMMSE}	= \m{W}\mf{H}_s^{-1}, \label{equ:LMMSE_001}
\end{equation}
where $\m{W}$ represents a Wiener smoothing matrix\footnote{Even though we use the same notation the Wiener smoothing matrix has nothing to do with the matrix in \eqref{equ:ZF_034}.} given by
\begin{equation}
		\m{W} = \m{G}_s^H \left(\m{G}_s\m{G}_s^H + \frac{N \sigma_n^2}{\sigma_d^2} (\mf{H}_s^H\mf{H}_s)^{-1}	\right)^{-1}. 					                 \label{equ:LMMSE_003}
\end{equation}
\eqref{equ:LMMSE_001} allows the following interpretation of the LMMSE estimator's mode of operation: The LMMSE equalizer acts as a composition of a simple channel inversion stage (multiplication with $\mf{H}_s^{-1}$ as in \eqref{equ:ZF_003}) and a Wiener smoothing operation (multiplication with $\m{W}$). The Wiener smoothing operation exploits the correlations between subcarrier symbols which have been introduced by \eqref{equ:3} at the transmitter, and acts as a noise reduction operation on the subcarriers. For the equalizer in \eqref{equ:LMMSE_001}, an $(N_d+N_r)\times (N_d+N_r)$ matrix has to be inverted. By applying the matrix inversion lemma, it can easily be shown that the equalizer can equivalently be determined by
\begin{equation}
	\m{E}_\mathrm{LMMSE} = (\m{G}_s^H\mf{H}_s^H \mf{H}_s\m{G}_s + \frac{N \sigma_n^2}{\sigma_d^2}\m{I})^{-1} \m{G}_s^H\mf{H}_s^H.       			              \label{equ:LMMSE_005}
\end{equation}
\eqref{equ:LMMSE_005} shows strong similarities to the BLUE in \eqref{equ:ZF_008}. For $\sigma_n^2 = 0$ the expressions for the LMMSE equalizer and the BLUE coincide. Note that by using \eqref{equ:LMMSE_005} instead of \eqref{equ:LMMSE_001} for the LMMSE equalizer determination the matrix to be inverted only has the dimension $N_d \times N_d$. The error $\vef{e}=\vef{d}-\widehat{\vef{d}}$ has zero mean, and its covariance matrix is given by
\begin{equation}
	\m{C}_{\tilde{e}\tilde{e}} = N \sigma_n^2 (\m{G}_s^H\mf{H}_s^H \mf{H}_s\m{G}_s + \frac{N \sigma_n^2}{\sigma_d^2}\m{I})^{-1}.       			              \label{equ:LMMSE_006}
\end{equation}

\subsection{Complexity Optimized LMMSE Batch Equalizer} \label{sec:compl}

For the LMMSE equalizer a complexity reduced version can be derived similar as for the BLUE in Sec.~\ref{sec:BLUE_optimized}. By introducing the real diagonal matrices
\begin{eqnarray}
	\m{D}_1 &=& \mf{H}_{s,1}^H \mf{H}_{s,1}+\frac{N \sigma_n^2}{\sigma_d^2}\m{I}, \\
	\m{D}_2 &=& \mf{H}_{s,2}^H \mf{H}_{s,2},
\end{eqnarray}
the expression $(\m{G}_s^H\mf{H}_s^H \mf{H}_s\m{G}_s + \frac{N \sigma_n^2}{\sigma_d^2}\m{I})^{-1}$ appearing in
\eqref{equ:LMMSE_005} and in
\eqref{equ:LMMSE_006} can be written as
\begin{eqnarray}
	\lefteqn{(\m{G}_s^H\mf{H}_s^H \mf{H}_s\m{G}_s + \frac{N \sigma_n^2}{\sigma_d^2}\m{I})^{-1} = } \nonumber \\
	      & & \m{D}_1^{-1}-\m{D}_1^{-1}\m{T}^H(\m{T}\m{D}_1^{-1}\m{T}^H+\m{D}_2^{-1})^{-1}\m{T}\m{D}_1^{-1}.  \label{equ:compl_003}
\end{eqnarray}
The derivation widely coincides with the one in Sec.~\ref{sec:BLUE_optimized}.

\subsection{Complexity Optimized Sequential LMMSE Receiver}\label{sec:sequential}
In this section we derive a highly complexity optimized sequential LMMSE receiver. We address the equalizer determination procedure as well as the data estimation process. The sequential LMMSE estimator completely avoids matrix inversions. Again the preparatory steps described in section \ref{sec:uw} - especially the re-sorting in \eqref{equ:system_model_002} - are extremely beneficial for the derivation of the complexity optimized solution. In this section we use the system model
\begin{equation}
	\vef{y} = \mf{H}_s \vef{c}_s + \vef{v}, \label{equ:sequ_007}
\end{equation}
and estimate $\vef{c}_s$ which includes both the data and the redundant subcarrier symbols. It turns out that using the system model in \eqref{equ:sequ_007} instead of the one in \eqref{equ:system_model_001} drastically simplifies the sequential LMMSE procedure since $\mf{H}_s$ is diagonal, in contrast to $\mf{H}_s\m{G}_s$. We let $\widehat{\vef{c}}_s[n]$ be the LMMSE estimate based on the first $n+1$ elements $\{\tilde{y}[0], \tilde{y}[1],\ldots,\tilde{y}[n]\}$ of the vector $\vef{y}$, and $\breve{\m{M}}[n]$ be the corresponding minimum MSE matrix
\begin{equation}
	\breve{\m{M}}[n] = E[(\vef{c}_s-\widehat{\vef{c}}_s[n])(\vef{c}_s-\widehat{\vef{c}}_s[n])^H].
\end{equation}
Furthermore, $\vef{h}_s[n]$ denotes the column vector that corresponds to the Hermitian transpose of the $n^{th}$ row of $\mf{H}_s[n]$. The sequential LMMSE estimator for the Bayesian linear model as in \eqref{equ:sequ_007} becomes (cf.~\cite{Kay93}):\newline

\noindent \textbf{Initialization:}
\begin{eqnarray}
	\widehat{\vef{c}}_s[-1] &=& E[\vef{c}_s] = \ve{0}\\
	\breve{\m{M}}[-1]       &=& \m{C}_{\tilde{c}_s\tilde{c}_s}  \nonumber \\ 																																								&=& E\left[(\vef{c}_s-\widehat{\vef{c}}_s[-1])(\vef{c}_s-\widehat{\vef{c}}_s[-1])^H\right]
															\nonumber \\
													&=& \sigma_d^2\m{G}_s\m{G}_s^H = \sigma_d^2 \begin{bmatrix} \m{I} & \m{T}^H \\ \m{T} & \m{T}\m{T}^H 																		\end{bmatrix}.  \label{equ:sequ_011}
\end{eqnarray}

\smallskip\noindent For $n=0,1,\ldots,(N_d+N_r-1)$ do

\smallskip\noindent\textbf{Gain Vector Update:}

\begin{equation}
	\ve{k}[n] = \frac{\breve{\m{M}}[n-1]\vef{h}_s[n]}{\sigma_v^2 + \vef{h}_s^H[n]\breve{\m{M}}[n-1]\vef{h}_s[n]} 															  \label{equ:sequ_001}
\end{equation}

\bigskip\noindent \textbf{Minimum MSE Matrix Update:}

\begin{equation}
	\breve{\m{M}}[n] = (\m{I}-\ve{k}[n]\vef{h}_s^H[n])\breve{\m{M}}[n-1] \label{equ:sequ_002}
\end{equation}

\bigskip\noindent \textbf{Estimate Update:}

\begin{equation}
	\widehat{\vef{c}}_s[n] = \widehat{\vef{c}}_s[n-1]+\ve{k}[n](\tilde{y}[n]-\vef{h}_s^H[n]\widehat{\vef{c}}_s[n-1])                        \label{equ:sequ_003}
\end{equation}

\smallskip\noindent \eqref{equ:sequ_001} and \eqref{equ:sequ_002} can be regarded as the equalizer determination procedure that can completely be performed immediately after channel estimation. Note that only the final MSE matrix $\breve{\m{M}}[N_d+N_r-1]$ but all gain vectors ($\ve{k}[0],\ve{k}[1],\ldots,\ve{k}[N_d+N_r-1]$) are required to be stored until the next channel estimation update. \eqref{equ:sequ_003} describes the sequential data estimation procedure for one UW-OFDM symbol, that has to be applied to every received OFDM symbol. After $(N_d+N_r)$ iterations the vector
\begin{equation}
	\widehat{\vef{d}} = \begin{bmatrix} \m{I} & \m{0} \end{bmatrix} \widehat{\vef{c}}_s,
\end{equation}
that contains the first $N_d$ entries of $\widehat{\vef{c}}_s$, exactly corresponds to the data estimate obtained when applying the batch LMMSE equalizers \eqref{equ:LMMSE_001} or \eqref{equ:LMMSE_005}. Further, the upper left $N_d \times N_d$ sub-matrix of $\breve{\m{M}}[N_d+N_r-1]$
\begin{equation}
	\m{C}_{\tilde{e}\tilde{e}} = \begin{bmatrix} \m{I} & \m{0} \end{bmatrix} \breve{\m{M}}[N_d+N_r-1] \begin{bmatrix} \m{I}
																										\\ \m{0} \end{bmatrix}
\end{equation}
exactly corresponds to the error covariance matrix in \eqref{equ:LMMSE_006}. \bigskip

In the following we significantly simplify the sequential LMMSE procedure. We first exploit the fact that the system matrix $\mf{H}_s$ is diagonal. Let $[\mf{H}_s]_{nn}$ be the $n^{th}$ main diagonal element of $\mf{H}_s$, $[\breve{\m{M}}]_{nn}$ be the $n^{th}$ main diagonal element of $\breve{\m{M}}$, $\breve{\ve{m}}_n$ be the $n^{th}$ column of $\breve{\m{M}}$, and $\widehat{\widetilde{c}}_{s,n}$ be the $n^{th}$ element of $\widehat{\vef{c}}_s$. Then since $\mf{H}_s$ is diagonal the iteration steps can be simplified to:\newline

\noindent \textbf{Gain Vector Update:}

\begin{equation}
	\ve{k}[n] = \frac{[\mf{H}_s]_{nn}^{*}\breve{\ve{m}}_n[n-1]}{\sigma_v^2 + |[\mf{H}_s]_{nn}|^2[\breve{\m{M}}]_{nn}[n-1]} 											\label{equ:sequ_004}
\end{equation}

\bigskip\noindent \textbf{Minimum MSE Matrix Update:}

\begin{equation}
	\breve{\m{M}}[n] = \breve{\m{M}}[n-1] - [\mf{H}_s]_{nn} \ve{k}[n] \ve{m}_n^H[n-1] \label{equ:sequ_005}
\end{equation}

\bigskip\noindent \textbf{Estimate Update:}

\begin{equation}
	\widehat{\vef{c}}_s[n] = 																																																				\widehat{\vef{c}}_s[n-1]+\ve{k}[n](\tilde{y}[n]-[\mf{H}_s]_{nn}\cdot\widehat{\widetilde{c}}_{s,n}[n-1])                        		\label{equ:sequ_006}
\end{equation}

\smallskip \noindent We notice that in the MSE matrix update equation a full matrix multiplication simplifies to a (column $\times$ row) multiplication, in the gain vector update equation a matrix-vector multiplication simplifies to a vector-scalar multiplication, and in the estimate update equation a vector inner product simplifies to a scalar product. \smallskip

The equations for the first $N_d$ iteration steps can further significantly be simplified by exploiting the fact that the data symbols are mutually uncorrelated, i.e.\ the upper left $N_d \times N_d$ submatrix of $\breve{\m{M}}[-1]$ is diagonal (and real), cf.~\eqref{equ:sequ_011}. We partition the gain vector and the MSE matrix as
\begin{equation}
	\ve{k} = \begin{bmatrix} \ve{k}_d \\ \ve{k}_r \end{bmatrix};
	\breve{\m{M}} = \begin{bmatrix} \breve{\m{M}}_d & \breve{\m{M}}_{dr}^H \\ \breve{\m{M}}_{dr} & \breve{\m{M}}_r \end{bmatrix},
\end{equation}
where the indices '$_d$' and '$_r$' indicate, that the corresponding vector and matrix entries correspond to data and redundant subcarrier symbols, respectively. Furthermore, we split all update equations in separate equations for the data and the redundant subcarrier symbols. The following consequences of $\breve{\m{M}}_d[-1]=\sigma_d^2\m{I}$ can be exploited:
\begin{itemize}
\item For all $n=0,1,\ldots,(N_d-1)$ the gain vector $\ve{k}_d[n]$ is non-zero only at its $n^{th}$ entry, and can therefore be replaced by the scalar gain factor $k_d[n]$. Consequently, only one data symbol $\widehat{\widetilde{d}}_n$ will be updated at the $n^{th}$ iteration step, so the estimate update equation for the data entries simplifies to a scalar equation. Since the data estimates are initialized with zeros ($\widehat{\widetilde{d}}_n[-1] = 0$), the data estimate update equation becomes particularly simple.
\item For all $n=0,1,\ldots,(N_d-1)$ the matrix $\breve{\m{M}}_d[n]$ is diagonal and real, and at the $n^{th}$ iteration step (again for $n=0,1,\ldots,(N_d-1)$) it only needs to be updated at its $n^{th}$ main diagonal element $[\breve{\m{M}}_d]_{nn}[n]$. Consequently, the MSE matrix update equation for $\breve{\m{M}}_d[n]$ also reduces to a scalar equation.
\item Due to similar arguments the matrix update for $\breve{\m{M}}_{dr}[n]$ simplifies to an update of its $n^{th}$ column vector $\breve{\ve{m}}_{dr,n}[n]$ for $n=0,1,\ldots,(N_d-1)$.
\end{itemize}
The iteration equations for $n=0,1,\ldots,(N_d-1)$ finally simplify as follows:\newline

\noindent \textbf{Gain Vector Update:}

\begin{eqnarray}
	k_d[n] 			&=& \frac{[\mf{H}_s]_{nn}^{*}}{\frac{\sigma_v^2}{\sigma_d^2} + |[\mf{H}_s]_{nn}|^2} \label{equ:sequ_008} \\
	\ve{k}_r[n] &=& \frac{[\mf{H}_s]_{nn}^{*}\breve{\ve{m}}_{dr,n}[n-1]}{\sigma_v^2 + \sigma_d^2|[\mf{H}_s]_{nn}|^2} 																\label{equ:sequ_012}
\end{eqnarray}

\bigskip\noindent \textbf{Minimum MSE Matrix Update:}

\begin{align}
	[\breve{\m{M}}_d]_{nn}[n] &= \frac{\sigma_v^2}{\frac{\sigma_v^2}{\sigma_d^2} + |[\mf{H}_s]_{nn}|^2} \label{equ:sequ_009} \\
	\breve{\ve{m}}_{dr,n}[n] 	&= \breve{\ve{m}}_{dr,n}[n-1] -  \sigma_d^2 [\mf{H}_s]_{nn} \ve{k}_r[n]  \label{equ:sequ_013} \\
	\breve{\m{M}}_r[n] 				&= \breve{\m{M}}_r[n-1] - [\mf{H}_s]_{nn} \ve{k}_r[n] \ve{m}_{dr,n}^H[n-1] \label{equ:sequ_014}
\end{align}

\bigskip\noindent \textbf{Estimate Update:}

\begin{eqnarray}
	\widehat{d}_n 				&=& k_d[n]\tilde{y}[n]  \label{equ:sequ_010} \\
	\widehat{\vef{r}}[n] 	&=& \widehat{\vef{r}}[n-1]+\ve{k}_r[n]\tilde{y}[n]  \label{equ:sequ_015}
\end{eqnarray}

\smallskip\noindent For the first $N_d$ iteration steps the highly complexity reduced equations \eqref{equ:sequ_008} to \eqref{equ:sequ_014} can be used for the equalizer determination, only for the last $N_r$ iteration steps the more complex (but also quite simplified) equations \eqref{equ:sequ_004} and \eqref{equ:sequ_005} have to be evaluated. Similarly the first $N_d$ iteration steps of the data estimation procedure for an UW-OFDM symbol can be performed using the highly complexity reduced equations \eqref{equ:sequ_010} and \eqref{equ:sequ_015}, while for the last $N_r$ steps \eqref{equ:sequ_006} has to be used. The complexity analysis will be given in section \ref{sec:complexity}.\smallskip

The derived procedure also allows for a quite intuitive interpretation of the mode of operation of the sequential LMMSE estimator: For $n=0,1,\ldots,(N_d-1)$ only one data symbol is updated in each iteration step. Consequently, during the first $N_d$ iterations we only count one single complex multiplication per data subcarrier symbol as in classical CP-OFDM. Merely the redundant symbols (which require a vector update) are truly updated from step to step. Only for the last $N_r$ iteration steps also the data subcarrier symbols are updated from iteration to iteration by utilizing the correlation information contained in the redundant subcarrier symbols.\smallskip

Note that these simplifications would not have been possible without the re-sorting step in \eqref{equ:system_model_002}. Without the re-sorting step the gain vector would already be filled completely in a very early iteration step, namely immediately after the first redundant subcarrier symbol appears within $\vef{y}$. Furthermore, if we had used the system model \eqref{equ:system_model_001} instead of \eqref{equ:sequ_007}, then we would have to perform full $(N_d\times N_d)\cdot(N_d\times N_d)$ matrix multiplication operations for the last $N_r$ iteration steps of the MSE matrix update.

\section{Complexity analysis} \label{sec:complexity}

In this section we will analyze the computational complexity of the derived equalizers on the one hand, and of the corresponding data estimation procedures on the other hand. These investigations will clearly show the benefits of the complexity reduced versions. In practice, the equalizers need to be determined each time the channel estimate is updated.

\subsection{Prerequisites}
We are aware that it is difficult or even impossible to declare an equitable measure of complexity, since the complexity of an implementation strongly depends on the choice of the hardware and software architecture and of many implementation details. Some operations can even be implemented in many different ways, which might have advantages on certain architectures as well. To simplify things we basically count the number of complex multiplication equivalents (CME) for each individual equalizer and for the corresponding data estimation procedure. We completely ignore additions. Complex division are counted as $1$ CME. Since the number of required divisions is negligible, this simplification does not effect the final complexity considerably. Real multiplications and real divisions are counted as $\tfrac{1}{4}$ CME. \smallskip

For many of the derived equalizer implementations we have to deal with matrix products of the form $\m{A}^{-1}\m{B}$ with a positive definite Hermitian matrix $\m{A}\in\mathbb{C}^{m \times m}$ and with $\m{B}\in\mathbb{C}^{m \times n_b}$. We notice that calculating $\m{X}=\m{A}^{-1}\m{B}$
is equivalent to solving the systems of simultaneous linear equations
\begin{equation}
	\m{AX}=\m{B}. \label{eq:inv_system2}
\end{equation}
For our complexity calculations we assume that \eqref{eq:inv_system2} is solved with the help of a Cholesky decomposition of $\m{A}$ given by $\m{A} = \m{L}\m{L}^H$,
where $\m{L}$ is a lower triangular matrix having positive values on its main diagonal. $\m{AX}=\m{B}$ can be rewritten as $\m{L}(\m{L}^H\m{X})=\m{B}$. To obtain $\m{X}$ one can solve $\m{LY}=\m{B}$ for $\m{Y}$ with the help of a forward substitution, and subsequently solve $\m{L}^H\m{X}=\m{Y}$ for $\m{X}$ with the help of a backward substitution. The Cholesky decomposition requires $\tfrac{1}{6}m^3$ complex multiplications/divisions and $m$ square roots \cite{Golub96}, \cite{Schwab06}. We neglect the square roots and end up with $\tfrac{1}{6}m^3$ CME. A single forward or backward substitution requires $\tfrac{1}{2}m^2+\tfrac{1}{2}m$ CME.  To solve \eqref{eq:inv_system2} with the help of a Cholesky decomposition we can finally assume a total count of  $\tfrac{1}{6}m^3+m^2 n_b+m n_b$ CME.

\smallskip
Whenever possible, we take any simplifications into account, that a special matrix structure (e.g. a diagonal, a real or a Hermitian matrix) could offer. Exemplarily, if the result of a matrix product is Hermitian, e.g.\ as in $\m{X}=\m{A}^H\m{A}$, then only the main diagonal and the lower triangular part needs to be computed.

\subsection{Complexity of the Investigated Equalizers and Data Estimation Procedures}
Before performing the data estimation with the help of one of the investigated equalizers an OFDM symbol has to be transformed to frequency domain with a length-$N$-FFT (fast Fourier transform) which requires $\tfrac{1}{2}N \log_2(N)$ CME. Furthermore, as one of the preparatory steps the influence of the UW has to be subtracted as described in \eqref{equ:system_model_005}. Since we do not count additions/subtractions in our complexity considerations this step does not increase the CME count for the data estimation procedure.

\smallskip
In the following we consider the complexity of the equalizers investigated above. We start our complexity investigations with the most simple equalizer $\m{E}_\text{CI}$ as given in \eqref{equ:ZF_014}. To determine $\m{E}_\text{CI}$ only $N_d$ CME (namely complex divisions to invert $\mf{H}_{s,1}$) are required. The data estimation procedure for an OFDM symbol in frequency domain requires $N_d$ CME (namely complex multiplications).

\smallskip
To estimate the data part of an OFDM symbol with the help of $\m{E}_\text{TDW}$ one could first determine its matrix representation as in \eqref{equ:ZF_004}, and then estimate the data vector by performing the full matrix-vector product $\widehat{\vef{d}}=\m{E}_\text{TDW}\vef{y}$ which requires $N_d(N_d+N_r)$ operations. However, most of the individual operations required to perform the data estimation are trivial. The procedure starts with the multiplication $\mf{H}_s^{-1}\vef{y}$ ($N_d+N_r$ CME), next the permutation is applied and the zero subcarrier symbols are added (zero CME) in order to be able to transform back to time domain with a length-$N$-IFFT ($\tfrac{N}{2}\log_2(N)$ CME). In time domain a windowing takes place, where the guard interval samples are forced to zero (zero CME). Next a transformation back to frequency domain is performed ($\tfrac{N}{2}\log_2(N)$ CME), the zero subcarriers are excluded again, a re-sorting is done, and finally the data symbols are extracted (zero CME). So in total 
the data estimation procedure per OFDM symbol requires $N\log_2(N)+N_d+N_r$ CME. For the equalizer determination only $\mf{H}_s$ needs to be inverted which requires $N_d+N_r$ CME (namely complex divisions).

\smallskip
Next we investigate the complexity of the different BLUE and LMMSE estimator batch representations. For all implementations the data vector estimation for one OFDM symbol requires a full matrix vector product $\widehat{\vef{d}}=\m{E}\vef{y}$ with $N_d(N_d+N_r)$ CME. The complexity of the equalizer determination differs significantly for the different implementations. We start with the representation of the BLUE as in \eqref{equ:ZF_008} and with the LMMSE estimator as in \eqref{equ:LMMSE_005}. These two expressions merely differ in the regularization term which only adds a single arithmetic operation. We neglect this single operation and treat \eqref{equ:ZF_008} and \eqref{equ:LMMSE_005} as equally complex. Using \eqref{equ:compl_004} it is easy to see that the matrix multiplication $\m{X}_1=\mf{H}_s\m{G}_s$ only requires $N_d N_r$ CME. For the product $\m{X}_2=\m{X}_1^H\m{X}_1$ we can use the findings from \eqref{equ:compl_001}, namely $\m{X}_2=\mf{H}_{s,1}^H \mf{H}_{s,1} +\m{T}^H \mf{H}_{s,2}^H \mf{H}_{s,2} 
\m{T}$. By additionally exploiting the fact that $\m{X}_2$ is Hermitian, one can easily find that the matrix product $\m{X}_2=\m{X}_1^H\m{X}_1$ requires $\tfrac{1}{2}N_d^2N_r+N_dN_r+N_d+N_r$ CME.
Finally the operation $(\m{X}_2)^{-1}\m{X}_1^H$ requires $\tfrac{7}{6}N_d^3+N_d^2N_r+N_d^2+N_dN_r$ CME by using the Cholesky decomposition together with the forward and backward substitutions as mentioned above. The overall CME count for the BLUE in \eqref{equ:ZF_008} and the LMMSE estimator in \eqref{equ:LMMSE_005} therefore adds up to
\begin{multline}
	\tfrac{7}{6}N_d^3+\tfrac{3}{2}N_d^2N_r+3N_dN_r+N_d^2+N_d+N_r \text{   CME}.
\end{multline}
With similar considerations one can show that the LMMSE equalizer as expressed in \eqref{equ:LMMSE_001} requires
\begin{multline}
	\tfrac{7}{6}N_d^3+\tfrac{5}{2}N_d^2N_r+2N_dN_r^2+\tfrac{1}{6}N_r^3\\
	+N_d^2+\tfrac{3}{2}N_dN_r+\tfrac{5}{2}N_d+\tfrac{5}{2}N_r \text{   CME}.
\end{multline}
\smallskip
For the complexity optimized batch representations of the BLUE and the LMMSE estimator one has to determine the expressions in  \eqref{equ:compl_002} and \eqref{equ:compl_003}, respectively, followed by a matrix multiplication with $\m{G}_s^H\mf{H}_s^H$.
The simple inverses $\m{D}_1^{-1}$ and $\m{D}_2^{-1}$ require $\tfrac{5}{4}N_d+\tfrac{5}{4}N_r$ CME, to determine $\m{D}_1^{-1}\m{T}^H$ another $\tfrac{1}{2}N_dN_r$, and for $\m{T}[\m{D}_1^{-1}\m{T}^H]$ additional $\tfrac{1}{2}N_dN_r^2+\tfrac{1}{2}N_dN_r$ CME are required. The operation $(\cdot)^{-1} \m{TD}_1^{-1}$ demands $\tfrac{1}{6}N_r^3+N_dN_r^2+N_dN_r$ CME, and the multiplication with $\m{D}_1^{-1}\m{T}^H$ adds $N_d^2N_r$ CME. The determination of $\m{G}_s^H\mf{H}_s^H$ and the final multiplication add $N_dN_r+N_d^2N_r+N_d^2$ CME, cf. \eqref{equ:compl_004}, which totals to
\begin{multline}
	\tfrac{1}{6}N_r^3+2N_d^2N_r+\tfrac{3}{2}N_dN_r^2 \\
	+N_d^2+3N_dN_r+\tfrac{5}{4}N_d+\tfrac{5}{4}N_r  \text{ CME}.
\end{multline}

\smallskip
Finally, we investigate the complexity of the sequential LMMSE estimator regarded in section \ref{sec:sequential}. We start with the data vector estimation of one OFDM symbol. For the first $N_d$ iteration steps the estimate updates are performed using \eqref{equ:sequ_010} and \eqref{equ:sequ_015} which in total requires $N_dN_r+N_d$ CME (complex multiplications). For the last $N_r$ iteration steps \eqref{equ:sequ_006} has to be evaluated which in total requires $N_dN_r+N_r^2+N_r$ CME (complex multiplications). Consequently the equalization of one OFDM symbol requires
\begin{equation}
	2N_dN_r + N_r^2 + N_d + N_r \text{ CME}.
\end{equation}
For the equalizer determination we have to count the operations required for the gain factor updates and for the MSE matrix updates. For the first $N_d$ iteration steps the gain factor updates are performed using \eqref{equ:sequ_008} and \eqref{equ:sequ_012} which in total requires $N_d N_r+N_d$ complex multiplications and $2N_d$ real divisions (which we count as $\tfrac{1}{2}N_d$ CME). For the last $N_r$ iteration steps \eqref{equ:sequ_004} has to be evaluated which in total requires $N_d N_r + N_r^2+N_r$ complex multiplications, $N_r$ real multiplications ($\tfrac{1}{4}N_r$ CME) and $2N_r$ real divisions ($\tfrac{1}{2}N_r$ CME). The MSE matrix updates for the first $N_d$ iterations are performed using \eqref{equ:sequ_009} to \eqref{equ:sequ_014} which in total requires $N_d N_r^2 + N_d N_r$ complex multiplications and $N_d$ real divisions ($\tfrac{1}{4}N_d$ CME). For the last $N_r$ iteration steps \eqref{equ:sequ_005} has to be evaluated which in total requires $N_d^2N_r+2N_dN_r^2+N_r^3+N_dN_r+N_r^2$ 
complex multiplications. We finally arrive at
\begin{multline}
	N_d^2N_r+3N_dN_r^2+N_r^3 \\
	+4N_dN_r+2N_r^2+\tfrac{7}{4}N_d+\tfrac{7}{4}N_r \text{ CME}.
\end{multline}

\subsection{Numerical Example}
In the simulation section we will show results for a particular parameter setup. The most important parameters can be found in Tab. \ref{tab2}. For the complexity considerations only $N_d,N_r$ and $N$ are important. The particular choices in Sec. \ref{sec:sim} are $N_d=36, N_r=16,N=64$. Tab. \ref{tab:complexity} compares the complexity of the different equalizer representations and data estimation procedures, respectively, for that particular parameter setup. Note that for the data vector estimation per OFDM symbol we count the contribution of the FFT ($\tfrac{N}{2}\log_2(N)$ CME) which is required in all cases, and the additional effort contributed by the particular equalization procedure.

\begin{table}[htb]
\caption{Computational complexity of the introduced equalizers and data estimators.}
\centering
\begin{tabular}{|l|c|c|} \hline
											 																										& CME for equalizer  & CME for data est.  \\
	Equalization method																											& determination			& per OFDM symbol \\ \hline \hline
	$\m{E}_\text{CI}$ \eqref{equ:ZF_014} 																				& 36 								&	228	    				\\ \hline
	$\m{E}_\text{TDW}$ \eqref{equ:ZF_004} 																			& 52 								& 628							\\ \hline
	$\m{E}_\text{LMMSE}$ \eqref{equ:LMMSE_001} 																		& 127677 					& 2064						\\ \hline
	$\m{E}_\text{BLUE}$,$\m{E}_\text{LMMSE}$ \eqref{equ:ZF_008},\eqref{equ:LMMSE_005}	& 88612 			& 2064						\\ \hline
	$\m{E}_\text{BLUE}$,$\m{E}_\text{LMMSE}$ \eqref{equ:compl_002},\eqref{equ:compl_003} 	& 59068 	& 2064						\\ \hline
	Sequential LMMSE																												& 55387									& 1652						\\ \hline
\end{tabular}
\label{tab:complexity}
\end{table}

We observe, that the simple equalizers $\m{E}_\text{CI}$ and $\m{E}_\text{TDW}$ show a significantly lower complexity for the equalizer determination as well as for the data estimation per OFDM symbol. Concerning the BLUE and the LMMSE estimator we can state that the complexity optimized batch solutions reduce the equalizer determination complexity by around 33\% compared to the straightforward implementations in \eqref{equ:ZF_008} and \eqref{equ:LMMSE_005}, respectively. The complexity optimized sequential LMMSE estimator which completely avoids matrix inversions further reduces the equalizer determination complexity by another 6\%, and interestingly enough also the data estimation complexity can be reduced by 20\%.

\section{Simulation Results} \label{sec:sim}

In this section we evaluate the introduced receiver concepts in terms of their BER performance. We notice that all derived variants of an estimator (e.g. of the LMMSE estimator) perform equivalently. Different performance of distinct versions of an estimator would only be expected if fixed point implementations were regarded which is not the focus of our investigations.

\subsection{Simulation Setup}
We show simulation results with and without outer channel coding. For the case when an outer channel code is used, the block diagram in Fig.~\ref{fig:block_diagram_ISIT} is extended by an outer channel encoder and an interleaver at the transmitter side, and by a deinterleaver and decoder at the receiver side. We used the same outer convolutional encoder with the industry standard rate 1/2, constraint length 7 code with generator polynomials (133, 171) as defined in \cite{IEEE99}. A soft decision Viterbi algorithm is applied for decoding. The main diagonal of the appropriate matrix $\m{C}_{\tilde{e}\tilde{e}}$ is used to specify the varying noise variances along the data symbols after data estimation. We assumed perfect channel knowledge in the simulations to be presented below.   \smallskip

In \cite{Huemer10_1} we compared our UW-OFDM approach with the CP-OFDM based IEEE 802.11a WLAN standard \cite{IEEE99} and showed that UW-OFDM outperforms CP-OFDM in frequency selective indoor environments. In this work we use the same parameter setup as in \cite{Huemer10_1} which has been adapted to the 802.11a standard wherever possible. The most important parameters are specified in Table \ref{tab2}.
\renewcommand{\arraystretch}{1.2}
\begin{table} [htb]
\caption{\label{tab2} Main PHY parameters of the investigated UW-OFDM system.}
\begin{center}
\begin{tabular}  {|l|c|}  											\hline
Modulation scheme						& QPSK 					\\ 	\hline
Coding rates (outer code)		& uncoded, 1/2 	\\ 	\hline
FFT length									& 64 						\\ 	\hline
Occupied subcarriers				& 52 						\\ 	\hline
Data subcarriers						&	36 						\\ 	\hline
Redundant subcarriers				& 16 						\\ 	\hline
DFT period									& 3.2 $\mu$s		\\ 	\hline
Guard duration							&  800 ns				\\ 	\hline
Total OFDM symbol duration	& 3.2 $\mu$s 		\\ 	\hline
Subcarrier spacing	& 312.5 kHz 						\\ 	\hline
\end{tabular}
\end{center}
\end{table}
The sampling frequency has been chosen to be $f_s = 20~\mathrm{MHz}$. As in \cite{IEEE99} the indices of the zero subcarriers within an OFDM symbol $\vef{x}$ are set to \{0, 27, 28,...,37\}. The indices of the redundant subcarriers are chosen to be \{2, 6, 10, 14, 17, 21, 24, 26, 38, 40, 43, 47, 50, 54, 58, 62\}. This set (which can also be expressed by an appropriate permutation matrix $\m{P}$) minimizes the cost function in \eqref{equ:uw001}, and therefore also the mean energy of the redundant subcarriers. Since we focus on data estimation procedures in this work rather than on synchronization or channel estimation approaches we chose the zero UW for the BER simulations below. Note that in conventional CP-OFDM like in the WLAN
standard, the total length of an OFDM symbol is given by $T_{DFT} + T_{GI}$. However, the guard interval is part of the DFT
period in the UW-OFDM approach which leads to significantly shorter total symbol durations. Hence, the compared systems
show almost identical bandwidth efficiencies.\smallskip

\subsection{Simulation Results in the AWGN Channel}

\begin{figure}[!ht]
\centering
\includegraphics[trim=30 0 40 0, clip, width=3.5in]{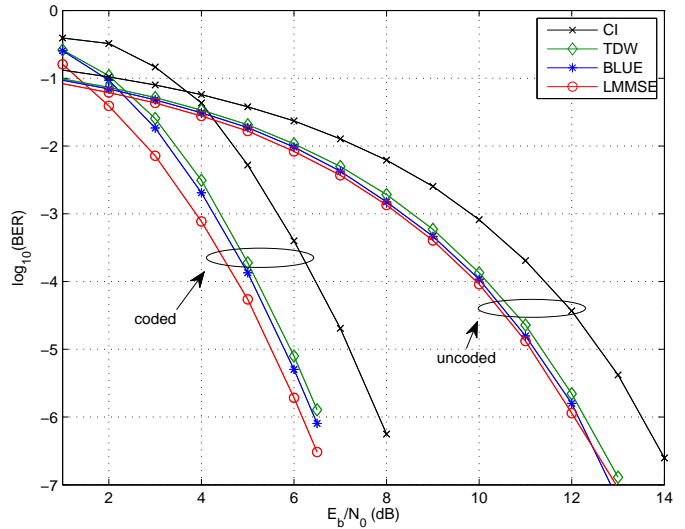}
\caption{Simulated BER performance of the investigated data estimators in the AWGN channel.}
\label{fig:BER_AWGN}
\end{figure}

Clearly, OFDM is designed for data transmission in frequency selective environments. Nevertheless, we start our comparison with simulation results in the AWGN channel, since these results provide first interesting insights. In Fig.~\ref{fig:BER_AWGN} the BER performance of the different data estimators is compared under AWGN conditions. As in all following BER figures we present curves for the case no outer code is used (we label it `uncoded' in the figures), and for an outer coding rate $r=\tfrac{1}{2}$.

We start the discussion with the uncoded case: As expected the CI estimator shows the worst performance, since it completely ignores the information present on the redundant subcarriers. Surprisingly, the very simple and intuitive TDW data estimator performs almost as well as the BLUE and the LMMSE in the AWGN environment. At a BER of $10^{-6}$ these three estimators which all make use of the a-priori knowledge introduced by the zero UW outperform the CI estimator by around 1.5dB. The trend is similar for $r=\tfrac{1}{2}$. However, it is completely in contrast to single carrier systems (e.g.\ SC/FDE) that the LMMSE estimator and the BLUE show a different performance in an AWGN environment. This comes from the fact that in single carrier systems the received QAM symbols are uncorrelated in an AWGN environment, whereas in UW-OFDM systems correlations are inherently present due to \eqref{equ:3}. However, the performance gain of the LMMSE estimator is quite small, and the BLUE approaches the LMMSE estimator 
performance for high $E_b/N_0$, as the term $\frac{N\sigma_n^2}{\sigma_d^2}$ in \eqref{equ:LMMSE_005} converges to zero.

\subsection{Simulation Results in Frequency Selective Indoor Environments}

\begin{figure}[ht]
\centering
\includegraphics[width=3.5in]{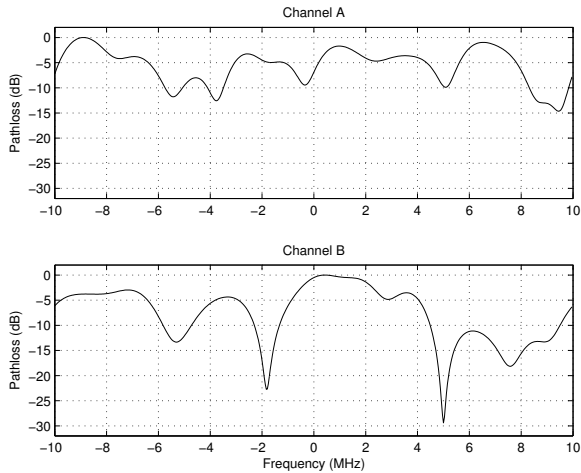}
\caption{Frequency responses of indoor multipath channel snapshots.}
\label{fig:channels}
\end{figure}

For the simulation of indoor multipath channels we applied the model described in \cite{Fak97}, which has also been used during the IEEE 802.11a standardization process. The channel impulse responses are modeled as tapped delay lines, each tap with uniformly distributed phase and Rayleigh distributed magnitude, and with power decaying exponentially. The model allows the choice of the channel delay spread. For a more detailed description we refer to \cite{Fak97}. For illustration purposes we use two different channel snapshots in this section, each channel featuring a delay spread of 100~ns, and a total duration not exceeding the guard interval. The frequency responses are shown in Fig.~\ref{fig:channels}.
Channel~A does not show any deep fading holes, whereas channel~B features two spectral notches within the system bandwidth, one at a data subcarrier position, the other one at a redundant subcarrier position.

\begin{figure}[t]
\centering
\includegraphics[trim=30 0 40 0, clip, width=3.5in]{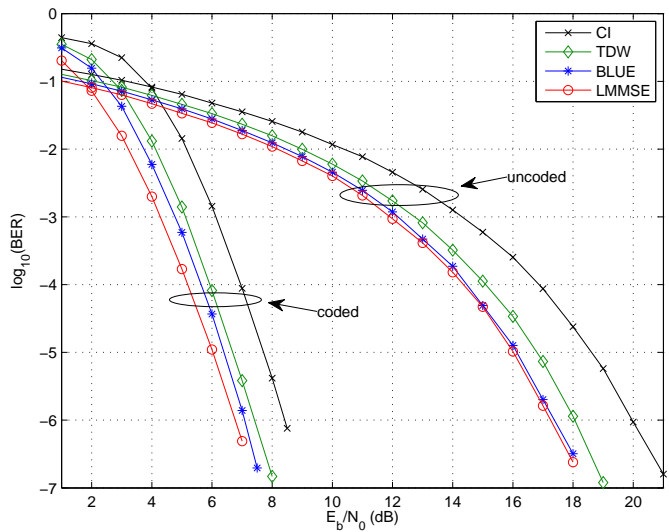}
\caption{Simulated BER performance of the investigated data estimators for channel A.}
\label{fig:BER_chA}
\end{figure}

Let us first interpret the results for channel~A, cf.\ Fig.~\ref{fig:BER_chA}. We observe similar trends as in the AWGN case, but now the LMMSE estimator and the BLUE clearly outperform the TDW estimator. For uncoded transmission the TDW outperforms the CI estimator by 1.9dB (again at a BER of $10^{-6}$), the BLUE and the LMMSE estimator gain 2.6dB and 2.7dB, respectively. For $r=\tfrac{1}{2}$ the corresponding gains shrink to 1.0dB, 1.35dB and 1.65dB, respectively. \smallskip

\begin{figure}[t]
\centering
\includegraphics[trim=30 0 40 0, clip, width=3.5in]{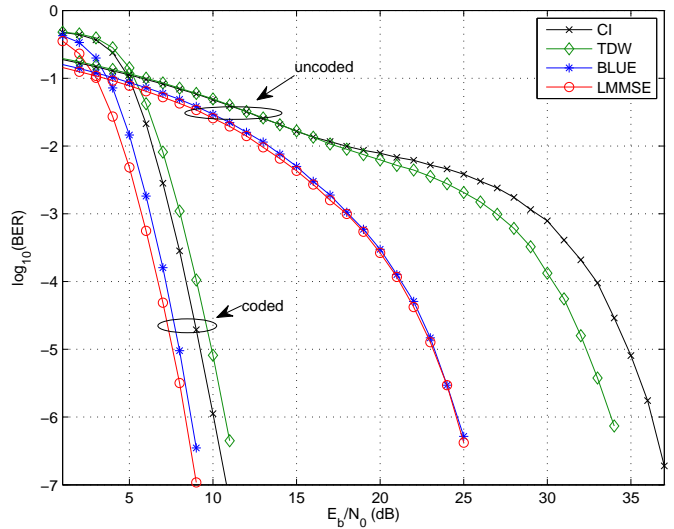}
\caption{Simulated BER performance of the investigated data estimators for channel B.}
\label{fig:BER_chB}
\end{figure}

Finally Fig.~\ref{fig:BER_chB} shows the simulation results for channel~B with its deep spectral notches. Very noticeable in the uncoded transmission is the bad performance of the CI and the TDW estimators. Here the performance gain of the BLUE and the LMMSE estimator is significant. The performance of the CI estimator is dominated by the weak BER behavior of data subcarrier symbols corresponding to deep spectral notches in the channel frequency response, while the LMMSE estimator (and similarly the BLUE) considerably decrease the noise on that subcarriers. (They decrease the noise variance on all subcarriers, but the effect is significant on subcarriers corresponding to deep spectral notches, cf. \cite{Huemer10_1}). The BLUE and the LMMSE estimator perform almost equivalently, which is again in contrast to SC/FDE systems, where the performance gain of the LMMSE estimator over the BLUE in channels with deep fading holes is usually much larger, particularly at low $E_b/N_0$ values, cf.~\cite{Huemer99b}. In 
coded transmission the performance loss of the CI estimator compared to the best performing LMMSE estimator decreases to 1.7dB. The significant improvement of the CI estimator in the coded case was expected as this corresponds to the usual coding gain as it is also observed in CP-OFDM. Somewhat unexpected, and in contrast to the uncoded results and those in an AWGN channel and in channel A, the TDW equalizer performs almost 0.7dB worse compared to the CI equalizer at a BER of $10^{-6}$. To understand this effect we will now have a closer look on the way the TDW estimator works. In fact, although it is hardly noticeable in Fig.~\ref{fig:BER_chB}, in the uncoded case the TDW only outperforms the CI estimator in the high $E_b/N_0$ range, but performs worse in the low $E_b/N_0$ range (0--15dB). However, this is the interesting $E_b/N_0$ range for coded transmission. We will now have a look on the noise variances (after equalization) and later on the BERs on the individual data subcarriers.

\begin{figure}[tbh]
\centering
\subfloat[Noise variances, total view]{
\includegraphics[width=3.43in, clip, trim=3 3 0 13]{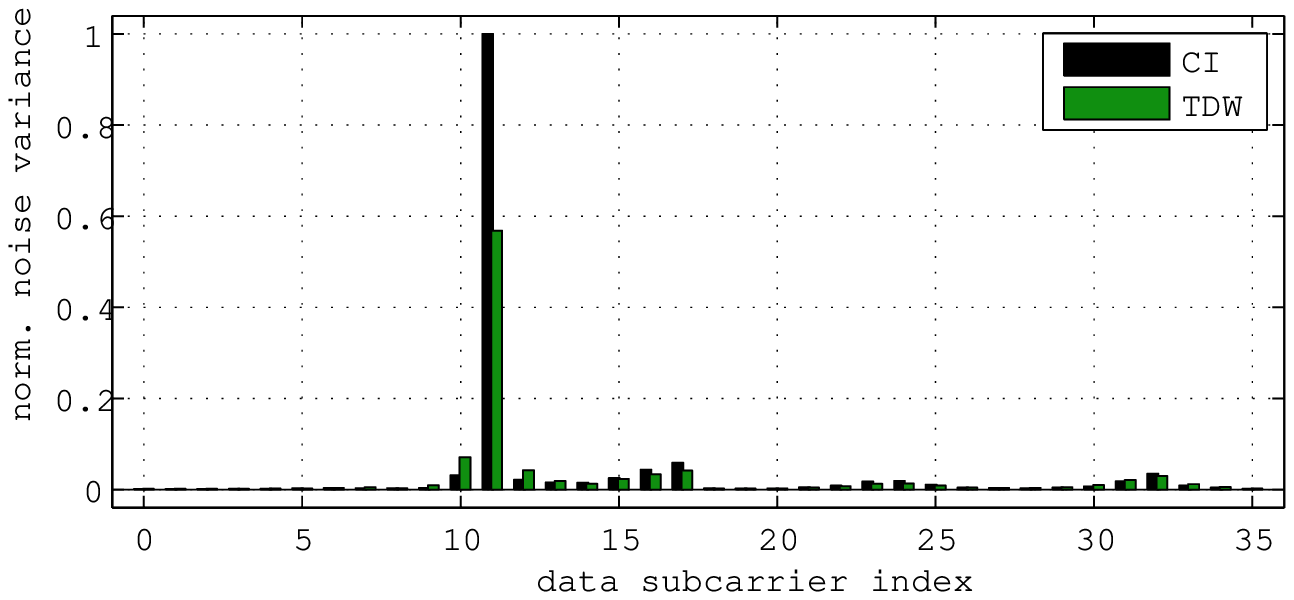}
\label{fig:noise_variances_total}
} \\
\subfloat[Noise variances, zoomed in]{
\includegraphics[width=3.5in, clip, trim=7 2 2 4]{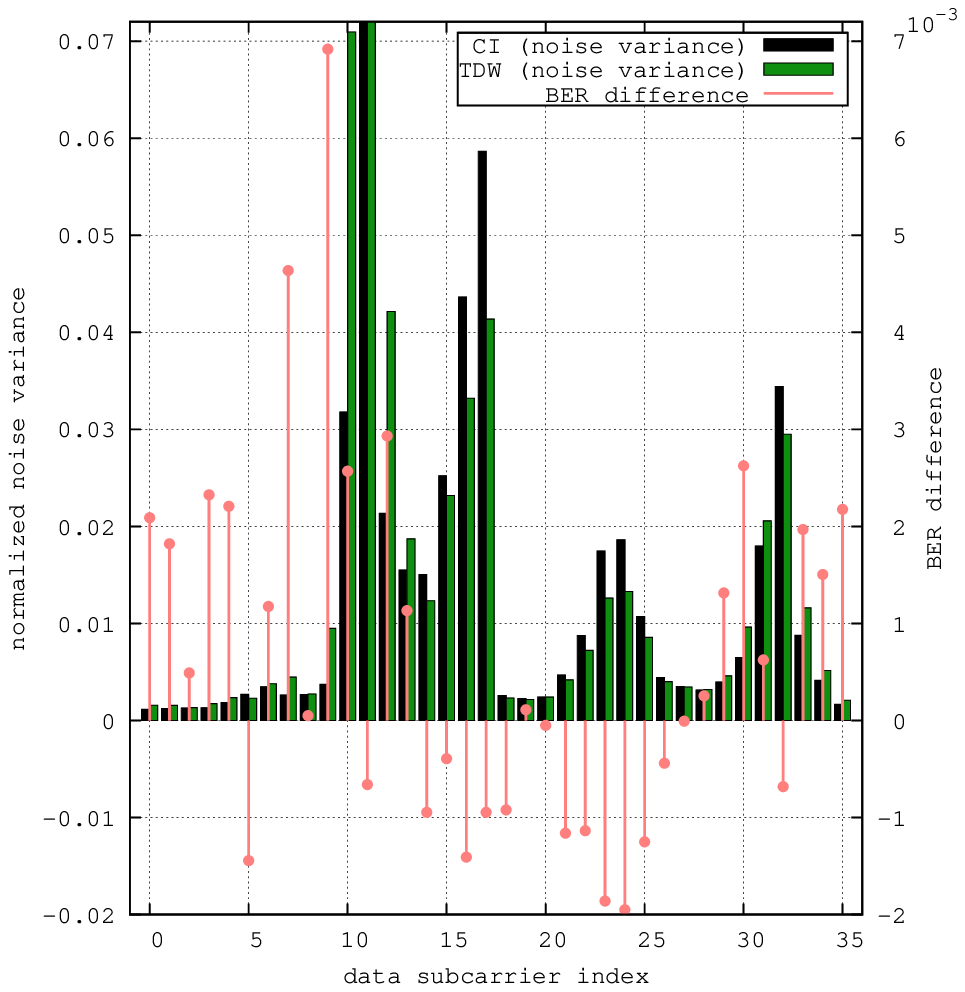}
\label{fig:noise_variances_zoom}
}
\caption{Subchannel noise variances after CI and TDW data estimation, and BER difference per subchannel.}
\label{fig:noise_variances}
\end{figure}

Fig.~\ref{fig:noise_variances_total} and \ref{fig:noise_variances_zoom} show the normalized noise variances after equalization at a fixed $E_b/N_0$ ($E_b/N_0=4~\mathrm{dB}$) for both data estimators. We observe that on the data subcarrier with index 11 the noise variance is tremendously reduced by the TDW compared to the CI estimator. This data subcarrier corresponds to the deep spectral notch around 5~MHz in the channel's frequency response. However, we also notice that the noise variances on data subcarriers around data symbol No. 11 are a little bit higher for the TDW compared to the CI estimator. On average (when averaged over all data subcarriers) the TDW equalizer clearly reduces the noise power compared to the CI equalizer, but besides a significant noise reduction on highly attenuated subcarriers, the TDW equalizer `distributes' some noise onto neighboring subcarriers.
Fig.~\ref{fig:noise_variances_zoom} additionally shows the difference between the resulting BERs of the TDW and the CI estimators on a subcarrier basis. We observe, that the tremendous noise reduction by the TDW equalizer on the 11$^{th}$ data subcarrier indeed leads to a lower subcarrier BER compared to the CI equalizer, but the improvement is minor. In return, the higher noise variances on the adjacent data subcarriers lead to increased corresponding subcarrier BERs for the TDW estimator. In total the increase of these subcarrier BERs lead to a worse overall BER performance of the TDW compared to the CI estimator for these $E_b/N_0$ values. The overall noise reduction by the TDW estimator is not translated to an overall BER gain for that particular channel for these $E_b/N_0$ values.


\section{Conclusion}
\label{sec:conclusion}

In this work we investigated several linear data estimators specifically designed for UW-OFDM. We introduced data estimators following the principles of classical estimation theory which lead to ZF equalizers. Two simple and intuitive ZF equalizers and the optimum ZF equalizer corresponding to the BLUE have been discussed. Following the Bayesian estimation principle the LMMSE estimator has been presented, and its batch and sequential versions have been regarded. We derived highly complexity reduced versions of the individual estimators and investigated their complexity in detail in terms of equivalent complex multilication counts. The CME count of the complexity optimized BLUE and LMMSE estimator versions could considerably be reduced compared to their straightforward counterparts, but still they show a significantly higher CME count compared to the simple ZF solutions. With the help of simulations we demonstrated the bit error behavior of the proposed estimators in the AWGN channel and in frequency 
selective indoor environments. Especially in frequency selective channels featuring deep fading holes the BLUE and in particular the LMMSE estimator significantly outperform the simple ZF estimators.

\section*{Acknowledgment}
\noindent
The authors want to express their deep thanks to the anonymous reviewers for many valuable comments.

\begin{IEEEbiography}[{\includegraphics[width=1in,height=1.25in,clip,keepaspectratio]{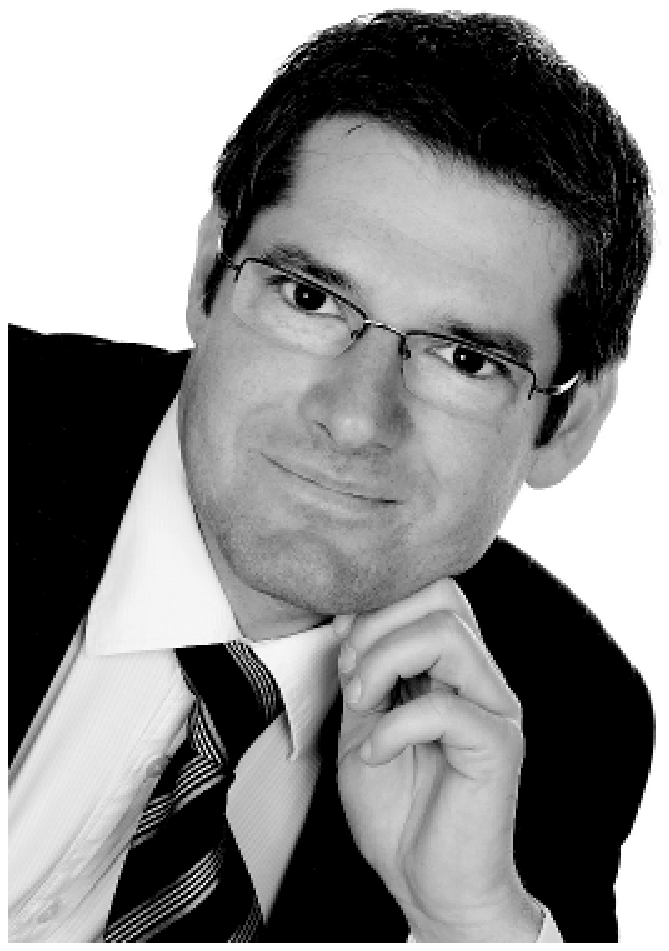}}]{Mario Huemer}
was born in Wels, Austria in 1970. He received the Dipl.-Ing. degree in mechatronics and the Dr.techn. (Ph.D.) degree from the Johannes Kepler University of Linz, Austria, in 1996 and 1999, respectively. From 1997 to 2000, he was a scientific assistant at the Institute for Communications and Information Engineering at the University of Linz, Austria. From 2000 to 2002, he was with Infineon Technologies Austria, research and development center for wireless products. From 2002-2004 he was a Professor for Communications and Information Engineering at the University of Applied Sciences of Upper Austria, from 2004-2007 he was Associate Professor for Electronics Engineering at the University of Erlangen-Nuremberg, Germany. In 2007, he has moved to Klagenfurt, Austria, to overtake the Directorship of the Chair for Embedded Systems and Signal Processing at the University of Klagenfurt as a Full Professor.

He has been engaged in research and development on WLAN, wireless cellular, and wireless positioning systems, and on highly integrated baseband and RFICs for mobile devices. Within these fields he published more than 110 papers. His current research interests are focused on signal processing algorithms and architectures for various applications.

Dr. Huemer is member of the IEEE Signal Processing Society, IEEE Communications Society and the IEEE Circuits and Systems Society. He is also member of the European Association for Signal Processing (EURASIP), the German Information Technology Society (ITG) in the Association for Electrical, Electronic and Information Technologies (VDE), and the Austrian Electrotechnical Association (OVE).
\end{IEEEbiography}

\begin{IEEEbiography}[{\includegraphics[width=1in,height=1.25in,clip,keepaspectratio]{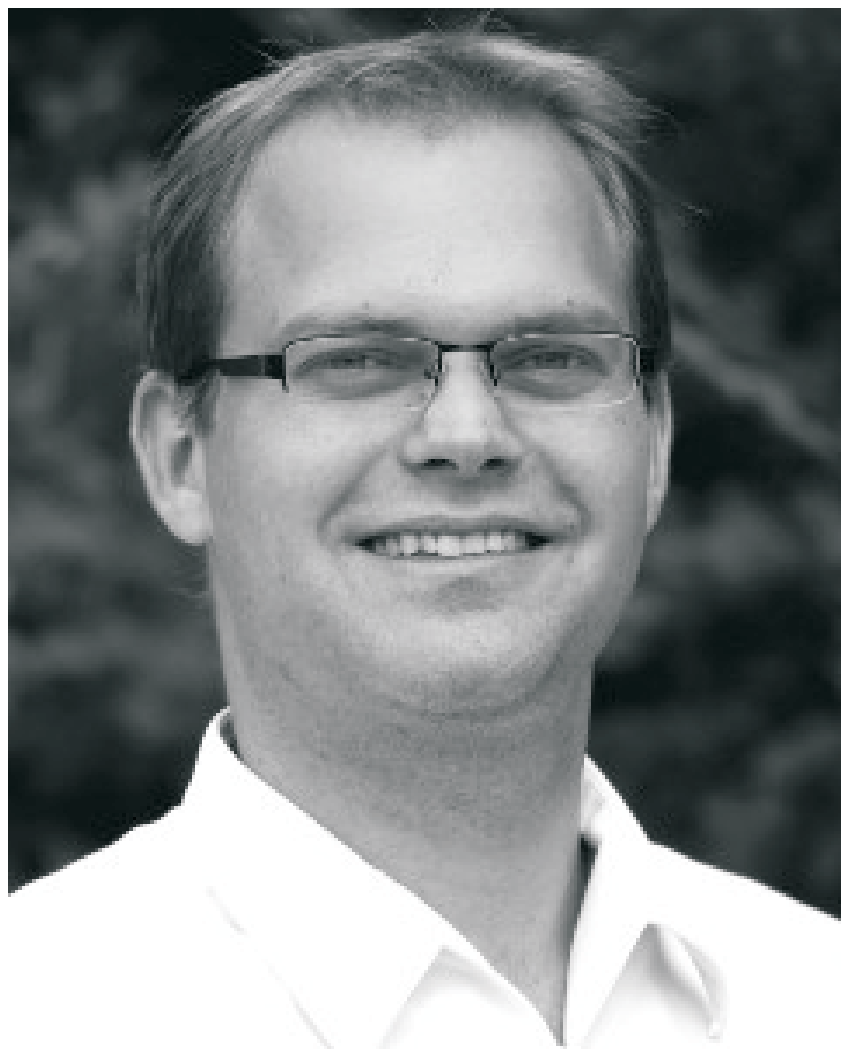}}]{Alexander Onic}
Alexander Onic was born in 1981. He started his studies of electrical engineering at the University of Erlangen-Nuremberg in 2001. After choosing information technology as his major subject he emphasized on signal processing and information theory in his education. He concluded the studies and received the Dipl.-Ing. degree in April 2007. The same month he joined the Embedded Systems and Signal Processing Group at Klagenfurt University.
\end{IEEEbiography}

\begin{IEEEbiography}[{\includegraphics[width=1in,height=1.25in,clip,keepaspectratio]{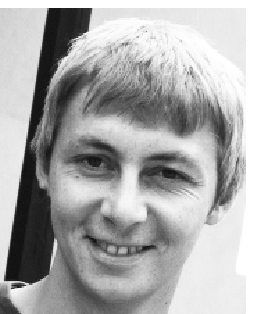}}]{Christian Hofbauer}
was born in 1982 in St. Peter am Wimberg, Austria. Between 2002 and 2006, he studied Hardware/Software Systems Engineering at the University of Applied Sciences of Upper Austria. He earned his DI (FH) degree, doing his thesis "High level refinement and implementation cost estimation for Matlab/Simulink models, targeting handheld Software Defined Radio architectures" at the research institute IMEC in Belgium. In January 2007, he started investigating spatial diversity techniques for MIMO systems at IMEC. Since September 2007, Christian is a member of the Embedded Systems and Signal Processing Group of Professor Dr. Mario Huemer. Currently, his research focuses on the investigation of the UW-OFDM concept.
\end{IEEEbiography}
\vfill

\end{document}